\documentclass[12pt]{article}
\usepackage[utf8]{inputenc}
\usepackage{graphicx,psfrag,epsf}
\usepackage{booktabs}
\usepackage{textgreek}
\usepackage{threeparttable}
\usepackage{float}
\usepackage{amsmath,amsfonts,amsthm,bm} % Math packages
\usepackage{url}
\usepackage{transparent}
\usepackage{adjustbox}
\usepackage{flafter}
\usepackage{lscape}
\usepackage{caption}
\usepackage{subcaption}
\usepackage{graphicx}
\usepackage{geometry}
\usepackage{footnote} % For footnotes in a table
\makesavenoteenv{tabular} % For footnotes in table
\usepackage{verbatim}
\usepackage{natbib}
\usepackage{comment}
\usepackage{rotating}
\usepackage{hyperref}
\usepackage{cleveref}
\usepackage{mdframed}
\usepackage{lipsum}
\usepackage{array}
\newcolumntype{H}{>{\setbox0=\hbox\bgroup}c<{\egroup}@{}}
\usepackage{xcolor}
\usepackage{amsmath}
\usepackage{multirow}
\usepackage{algorithmicx}
\usepackage{algorithm,algpseudocode}

\newcommand{\blind}{0}

\begin{document}

\def\spacingset#1{\renewcommand{\baselinestretch}%
{#1}\small\normalsize} \spacingset{1}

%----------------------------------------------------------------------------------------
%	Title Page Section
%----------------------------------------------------------------------------------------
\def\spacingset#1{\renewcommand{\baselinestretch}%
{#1}\small\normalsize} \spacingset{1}

%%%%%%%%%%%%%%%%%%%%%%%%%%%%%%%%%%%%%%%%%%%%%%%%%%%%%%%%%%%%%%%%%%%%%%%%%%%%%%

\if0\blind
{
  \title{\bf Nowcasting Growth using Google Trends Data: A Bayesian Structural Time Series Model\thanks{
   The authors thank Atanas Christev, Marco Del Negro, Sylvia Frühwirth-Schnatter, George Kapetanios, Sylvia Kaufmann, Gary Koop, Stephen Millard, Aubrey Poon, Galina Potjagailo, Mark Schaffer, Herman van Dijk, and all participants of the Annual Workshop on Financial Econometrics in Orebo (2019), the Bank of England Big Data Conference (2019), the CEF conference (2020), the International Symposium on Forecasting (2019,2020), the Panmure House PhD conference (2019),  the Study Center Gerzensee and Norges Bank conferences for their invaluable feedback. The constructive suggestions and comments from two anonymous reviewers and encouragement from the Editor helped us revise and improve upon the paper substantially. Their help is gratefully acknowledged. The usual disclaimer applies.}}
  \author{David Kohns\footnote{Corresponding author. Email: david.kohns94@googlemail.com} \hspace{.2cm}\\
    Economics, Heriot-Watt University, UK\\
    and \\
    Arnab Bhattacharjee \\
    Economics, Heriot-Watt University, UK \& \\
    National Institute of Economic \& Social Research, UK.}
  \maketitle
} \fi

\if1\blind
{
  \bigskip
  \bigskip
  \begin{center}
    {\LARGE\bf Title}
\end{center}
  \medskip
} \fi

%----------------------------------------------------------------------------------------
%	Abstract Section
%----------------------------------------------------------------------------------------
\begin{center}

\end{center}

\bigskip
\begin{abstract}
\noindent This paper investigates the benefits of internet search data in the form of Google Trends for nowcasting real U.S. GDP growth in real time through the lens of mixed frequency Bayesian Structural Time Series (BSTS) models. We augment and enhance both model and methodology to make these better amenable to nowcasting with large number of potential covariates. Specifically, we allow shrinking state variances towards zero to avoid overfitting, extend the SSVS (spike and slab variable selection) prior to the more flexible normal-inverse-gamma prior which stays agnostic about the underlying model size, as well as adapt the horseshoe prior to the BSTS. The application to nowcasting GDP growth as well as a simulation study demonstrate that the horseshoe prior BSTS improves markedly upon the SSVS and the original BSTS model with the largest gains in dense data-generating-processes. Our application also shows that a large dimensional set of search terms is able to improve nowcasts early in a specific quarter before other macroeconomic data become available. Search terms with high inclusion probability have good economic interpretation, reflecting leading signals of economic anxiety and wealth effects.

%applied to the original BSTS of \citet{scott2014predicting}. We rewrite the state space model into a non-centred representation which allows to shrink state variances to zero in order to avoid overfitting states. We further extend the SSVS prior on the regression coefficients to agnosticity of the underlying model size as well as compare its performance to the horseshoeprior of \citet{carvalho2010horseshoe}. To aid interpretability of the horseshoe prior, we also extend the SAVS algorithm of \citet{ray2018signal} on an iteration bases. The application to nowcasting GDP growth as well as a simulation study show that the horseshoe prior BSTS improves markedly over the SSVS and original BSTS model, with largest gains to be expected in dense designs.
\end{abstract}

%----------------------------------------------------------------------------------------
%	Key words
%----------------------------------------------------------------------------------------

\noindent%
\vfill

\newpage
\spacingset{1.45} % DON'T change the spacing!

%----------------------------------------------------------------------------------------

\section{Introduction} \label{sec:intro}
% 1. Motivation for Nowcasting
The primary object of nowcasting models is to produce `early' forecasts of target variables associated with long delays in data publication by exploiting the real time data publication schedule of the explanatory data set. While prediction is the primary goal here, the selected models can also sometimes provide structural interpretations \textit{ex post}. Nowcasting is particularly relevant to central banks and other policymakers who are tasked with conducting forward looking policies on the basis of key economic variables such as GDP or inflation. Inflation data are, however, published with a lag of up to 7 weeks with respect to their reference period, and precise estimates of GDP can take years.\footnote{The exact lag in publications of GDP and inflation depends as well on which vintage of data the econometrician wishes to forecast. Since early vintages of aggregate quantities such as GDP can display substantial variation between vintages, this is not a trivial issue.} Since even monthly macroeconomic data arrive with considerable lag, it is now common to combine, next to traditional macroeconomic data, ever more information from Big Data sources such as internet search terms, satellite data, scanner data, etc. which have the advantage of being available in near real time \citep{bok2018macroeconomic}. The recent Covid-19 pandemic has given further impetus to this trend, as faster indicators have proven especially useful in modelling the unprecedentedly sharp movements in the economy that traditional macroeconometric models fail to capture in a timely manner \citep{antolin2021advances,woloszko2020tracking}.

% 2. Motivation for Google Trends
%In this paper we add to the burgeoning literature on using Google search data in the form of Google Trends (GT), which measure the relative search volume of certain search terms entered into the Google search engine, to nowcast aggregate economic time-series. Although the majority of the Google Trends nowcasting literature has shown that there can be substantial gains using internet search information compared to simple autoregressive processes when the economic link between the search activity and the economic outcome is clear, such as in vehicles, labour market decisions, relatively little previous research has investigated the use of Google Trends to nowcast aggregate economic time series, to which a structural link can be more opaque. Notable exceptions....  [To motivate the potential usefulness of GT, for nowcasting GDP growth, consider figure X]. We contribute to this literature, by investigating the benefits of monthly Google search information for nowcasting quarterly US real GDP growth in real-time compared to traditional macro data and survey information.

In this paper we add to the burgeoning literature on using Google search data in the form of Google Trends (GT), which measure the relative search volume of certain search terms entered into the Google search engine, to nowcast aggregate economic time-series. In particular, we investigate the benefits of using monthly Google search information for nowcasting quarterly U.S. real GDP growth in real-time compared to traditional macro data and survey information. We contribute to this literature by being, to our best knowledge, the first paper to investigate the benefit of search information above and beyond macroeconomic data for the U.S. including performance during the Covid-19 pandemic period.

%While previous nowcasting applications of Google Trends are vast (see for example \citet{niesert2020can} or \citet{koop2019macroeconomic}), relatively few articles have investigated potential of Google Trends for nowcasting GDP growth. Notable exceptions are  \citet{ferrara2019google} who have found that Google Trends improve nowcasts of quarterly Euro Area growth prior to arrival of more traditional macroeconomic information and recently \citet{woloszko2020tracking} who builds a GDP growth tracker on the basis of a host of Google search information. To the best of our knowledge, we are the first paper to investigate the benefit of search information above and beyond macro data for the US which also investigates performance during the Covid-19 pandemic period.

To deal with the specificities of the data, we propose robust nowcasting methods that are amenable to situations in which the policymaker needs to combine traditional and non-traditional data sources while providing tractable variable selection properties. For this purpose, we adapt current generation state space and regression priors to the widely popular Bayesian Structural Time Series model (BSTS) \citep{scott2014predicting}. Results from our nowcasting application show that Google's search information improves nowcasts of GDP growth, particularly early on in the quarter before macroeconomic data are published. We show that our extensions allow for accuracy gains of up to 40\% during certain nowcast periods in point as well as in density nowcasts compared to the original BSTS model of \citet{scott2014predicting} while retaining its interpretability. These results are confirmed in a simulation study which checks robustness to a variety of data-generating processes.

% Overview Rest of paper
In the following, we firstly discuss the state and regression priors as well as posteriors for our extended BSTS models and provide efficient sampling algorithms. In section 3, we elaborate further on the data used for nowcasting, including dealing with mixed frequency, the data publication calendar and the specificities of the Google Trends data set. In section 4, we present results based on our empirical application of nowcasting U.S. GDP growth, which is followed in section 5 by the results from our simulation study. Finally, Section 6 concludes with a discussion and avenues for future research.

\section{Bayesian Structural Time Series} \label{sec:meth}
\subsection{The Original Model}
% 1. Original BSTS (Mention attractiveness for GDP in the intro)
The Bayesian Structural Time Series (BSTS) model, as proposed by \citet{scott2014predicting}, provides a conceptually attractive model for nowcasting  aggregate economic time-series with heterogeneous data sources, as it flexibly estimates latent time-trends, seasonality and deviations or `irregular' dynamics through variable selection using a high-dimensional shrinkage prior. Denote the target variable to be nowcast by $y_t = (y_1,\cdots,y_T)'$ and the K-dimensional explanatory data set as $x_t = (x_1',\cdots,x_T')'$ which for now are sampled at the common frequency, $t$. Then our model is as follows: 

\begin{equation}
    \begin{split}
    y_t & = \tau_t + x_t'\beta + \delta_t + \epsilon_t, \epsilon_t \sim N(0,\sigma_y^2) \\
    \tau_t & = \tau_{t-1} + \alpha_t + \epsilon^{\tau}_t, \epsilon^{\tau}_t \sim N(0,\sigma_{\tau}^2) \\
    \alpha_t & = \alpha_{t-1} + \epsilon^{\alpha}_t, \epsilon^{\alpha}_t \sim N(0,\sigma_{\alpha}^2) \\
    \delta_t & = -\sum^{S-1}_{s=1}\delta_{t-s} + \epsilon^{\delta}_t, \epsilon^{\delta}_t \sim N(0,\sigma_{\delta}^2).
    \end{split} \label{equ:bsts}
\end{equation}

(\ref{equ:bsts}) is a linear state space model with Gaussian errors and states $\{\tau_t,\alpha_t,\delta_t\}_{t=1}^T$ which capture long-run trends and $S$ seasonal components $\delta_t$. The deviation from $\tau_t$ describes variation from a long-run trend which, when applied to the level of GDP can be interpreted as the output gap \citep{watson1986univariate,grant2017bayesian}. $\alpha_t$ allows for a drift term in the trend which is often observed in stock variables such as in GDP, aggregate consumption and inflation \citep{grant2017bayesian,chan2017stochastic}.

Variable selection on the possibly high-dimensional $K\times 1$ response vector $\beta$ in the BSTS model of \citet{scott2014predicting} is done via a two component conjugate spike-and-slab prior. Estimation is standard \citep{george1993variable}, and states $\tau_t$, $\alpha_t$ and $\delta_t$ are estimated jointly via the forward
filtering backward sampling (FFBS) algorithm of \citet{durbin2002simple} based on the Kalman filter. This implementation relies on Normal-Inverse Gamma (N-IG) priors for the states and state variances for conditional conjugacy. While the BSTS model is a natural model for many time-series applications, we bring 3 important methodological innovations which make it more robust to overfitting trend estimation and variable selection with heterogeneous high dimensional data.  

\subsection{Model enhancements}
\subsubsection{Non-Centred Bayesian Structural Time Series}\label{sub:Non_centred}
% GDP growth interpretation
In line with previous nowcasting studies, this paper focuses on nowcasting GDP \emph{growth} rather than levels. However, two problems arise when applying model (\ref{equ:bsts}) directly to growth variables. As growth variables are often approximately stationary, conceptually, the inclusion of $\alpha_t$ implies that GDP growth follows a boundless drift for which there is little structural justification or empirical evidence. A non-drifting stochastic trend, on the other hand, has been shown to markedly improve nowcasts of GDP growth as shown in \cite{antolin2017tracking}, especially when the state variances are tightly controlled by priors such that the stochastic trend does not wander too wildly. This suggests that modeling time-variation is preferred over de-trending a priori.\footnote{Modelling an I(1) component in U.S. GDP growth is additionally consistent with Harvey's local-linear trend model \citep{harvey1985trends}, the \citet{hodrick1997postwar} filter and \citet{stock2012disentangling}.} The underlying rationale for this improvement is the well known empirical finding of changes in long-run GDP growth \citep{kim1999has,mcconnell2000output,jurado2015measuring}.  Econometrically, the additional problem is that the IG priors with no prior mass on zero, as implemented for Bayesian linear state space methods, can bias posterior state variances away from zero, thereby potential leading to false support for state dynamics which can hurt forecast performance.

We extend model (\ref{equ:bsts}) to flexibly let the data shut down state dynamics, and therefore broaden the applicability of model (\ref{equ:bsts}), by adopting the non-centred parameterisation of the state space as suggested by \citet{fruhwirth2010stochastic}. The non-centred parameterisation models state variances directly in the observation equation, which with normal priors, exerts much stronger shrinkage than IG priors.\footnote{Formally, a normal prior on the state standard deviation can be shown to imply a Gamma prior on the state variance.} This allows additionally for valid inference on testing for zero posterior variance via Savage-Dickey density ratios, as will be further discussed in section 5. Testing for zero posterior variance would be very challenging in a frequentist hypothesis testing approach because the null hypothesis of constant state in model (\ref{equ:bsts}) lies on the boundary of the parameter space.

The non-centred model considered for the empirical application is equivalently written as:
\begin{equation} \label{eq:non-centred1}
    y_t = \tau_0 + \sigma_{\tau}\tilde{\tau}_t + t\alpha_0 + \sigma_{\alpha}\sum_{s=1}^t\tilde{\alpha}_t + x_t'\beta + \epsilon_t, \; \epsilon_t \sim N(0,\sigma^2)
\end{equation}
and
\begin{equation} \label{eq:non-centred2}
    \begin{split}
        \tilde{\tau}_t & = \tilde{\tau}_{t-1} + \tilde{u}^{\tau}_t, \tilde{u}^{\tau}_t \sim N(0,1) \\
        \tilde{\alpha}_t & = \tilde{\alpha}_{t-1} + \tilde{u}^{\alpha}_t, \tilde{u}^{\alpha}_t \sim N(0,1)
    \end{split}
\end{equation} 
with starting values $\tilde{\tau}_0 = \tilde{\alpha}_0 = 0$. Note that the seasonal component is left out for estimation due to the small sample length of the Google Trends data set and differing seasonal patterns between monthly and quarterly data.\footnote{For further discussion, please see section \ref{sec:data}} To see that (\ref{eq:non-centred1}) and (\ref{eq:non-centred2}) is equivalent to (\ref{equ:bsts}), let:
\begin{equation}
    \begin{split}
        \alpha_t & = \alpha_0 + \sigma_{\alpha}\tilde{\alpha}_t \\
        \tau_t & = \tau_0 + \sigma_{\tau}\tilde{\tau}_t + t\alpha_0 + \sigma_{\alpha}\sum_{s=1}^t\alpha_s 
    \end{split}
\end{equation} \label{eq:non-centred3}
Hence, by setting $y_t = \tau_t + x_t'\beta + \epsilon_t$, it is clear that
\begin{equation}
    \begin{split}
        \alpha_t - \alpha_{t-1} & = \sigma{\alpha}(\tilde{\alpha}_t - \tilde{\alpha}_{t-1}) \\
        & = \sigma_{\alpha} + \tilde{u}^{\alpha}_t \\
        \tau_t - \tau_{t-1} & = \alpha_0 + \sigma_{\alpha}\tilde{\alpha}_t + \sigma_{\tau}(\tilde{\tau}_t - \tilde{\tau}_{t-1}) \\
        & = \alpha + \sigma_{\tau} + \tilde{u}^{\tau}_t
    \end{split}
\end{equation}
which recovers (\ref{equ:bsts}). Since $\sigma_{\tau,\alpha}$ are allowed to have support on the real line, they are not identified in multiplication with the states: the likelihood is invariant to signs of $\sigma_{\alpha}$ and $\sigma_{\tau}$. Consequently, mixing of the posterior state standard deviations can be poor and their distributions are likely to be bi-modal \citep{fruhwirth2010stochastic}. This issue is addressed by randomly permuting signs in the Gibbs sampler as explained below. Similar to \citet{fruhwirth2010stochastic}, we assume normal priors centred at 0 for $\sigma_i: \sigma_i \sim N(0,V_i)$ $\forall i \in \{\tau,\alpha\}$. 

Collecting all state space parameters in $\theta = (\tau_0,\alpha_0,\sigma_{\tau},\sigma_{\alpha})$, we assume an independent multivariate normal prior with diagonal covariance matrix:
\begin{equation}
    \theta \sim N(\theta_0,V_{\theta}).
\end{equation}
While the state processes $\{\tilde{\tau},\tilde{\alpha}\}^T_{t=1}$ can be estimated by any state space algorithm, we opt for the precision sampler method of \citet{chan2009efficient} which is outlined in Appendix (\ref{subsec:estimation}) along with the state posteriors. In contrast to FFBS type algorithms, it samples the states without recursive estimation which speeds up computation significantly.

%\section{Priors}
% 3: SSVS Prior
\subsubsection{SSVS Prior}
Our second enhancement concerns the SSVS prior. Variable selection in the BSTS model of \citet{scott2014predicting} is done via a two component conjugate spike-and-slab prior which utilises a variant of Zellner’s g-prior and fixed expected model size. While computationally fast due to conjugacy, many high-dimensional problems benefit from prior independence \citep{moran2018variance} and a fully hierarchical formulation to let the data decide on the most likely value of the parameters \citep{ishwaran2005spike}. 

Therefore, we follow \citet{ishwaran2005spike}'s extension to the SSVS prior, the Normal-Inverse-Gamma prior:
\begin{equation} \label{eq:nigprior}
    \begin{split}
    \beta_j|\gamma_j,\delta_j^2 & \sim \gamma_jN(0,\delta_j^2) + (1-\gamma_j)N(0,c\times\delta_j^2) \\
        \delta_j^2 & \sim Gamma(a_1.a_2) \\
        \gamma_j & \sim Bernoulli(\pi_0) \\
        \pi_0 & \sim Beta(b_1,b_2)
    \end{split}
\end{equation} 
where $j \in (1,\cdots,K)$. The intuition remains the same as compared to the spike-and-slab prior of \citet{scott2014predicting} in that the covariate's effect is modeled by a mixture of normals where it is either shrunk close to zero via a narrow distribution around zero (the spike component) or estimated freely though a relatively diffuse normal distribution (the slab component). Sorting into each component is handled through an indicator variable, $\gamma_j$, and the hyperparameter $c$ is chosen to be a very small number, thereby forcing shrinkage of noise variables to close to zero. While in the original BSTS model, the indicator variable, $\gamma_j$, depends on a fixed prior $\pi_0$ which governs the prior inclusion probability of a variable, (\ref{eq:nigprior}) allows for it to be estimated from the data through another level of hierarchy. We set $b_1=b_2=1$, which effectively assumes that any expected model size is a priori possible and thus allows for sparse but also dense model solutions as recommended by \cite{giannone2021economic}. Finally, the prior variance $\delta^2_j$ is also allowed to be hierarchical. Posteriors are standard and described in the Appendix (\ref{subsec:posteriors}). The posterior of $\gamma_j$ is of special interest to the analyst as it gives a data informed measure of importance of a variable. Specifically, $p(\gamma|y)$ can be interpreted as the posterior inclusion probability of a variable. \\
% 4. Horseshoe Prior
\subsubsection{Horseshoe Prior}
Our third and final enhancement of the BSTS models extends the employed shrinkage priors to the horseshoe prior. Like many recently popularised shrinkage priors, the horseshoe prior belongs to the broader class of global-local priors which take the following general form:
\begin{equation}
\begin{split}
     \beta_j | \lambda_j^2, \nu^2, \sigma^2 & \sim N(0,\lambda_j^2\nu^2\sigma^2), j \in (1,\cdots, K) \\
    \lambda_j & \sim \pi(\lambda_j)d\lambda_j, j \in (1,\cdots, K) \\
    \nu & \sim \pi(\nu)d\nu \\
    \sigma^2 & \sim \pi(\sigma^2)d\sigma^2 \\
    \end{split}
\end{equation} \label{eq:glpriors}
The idea of this family of priors is that the global scale $\nu$ controls the overall shrinkage applied to the regression, while the local scale $\lambda_j$ allows for the local possibility of regressors to escape shrinkage if they have large effects on the response. A variety of global-local shrinkage priors have been proposed \citep{polson2010shrink}, but here we focus on arguably the most popular, the horseshoe prior of \cite{carvalho2010horseshoe} which employs two half Cauchy distributions for $\nu$ and $\lambda_j$:
\begin{equation}  \label{prior5}
 \begin{split}
     \lambda_j & \sim C_+(0,1) \\
     \nu & \sim C_+(0,1)
 \end{split}
 \end{equation} 
These two fat tailed scale distributions imply a shrinkage profile that has the spike-and-slab prior in its limit and therefore offers a continuous approximation to the SSVS \citep{piironen2017sparsity} (see section \ref{robustness} in the appendix for further discussion). An additional attractive feature of the horseshoe prior is that it is completely automatic with respect to its hyperparameters and has been shown to be excellent at forecasting in several previous studies  \citep{huber2019inducing,huber2020nowcasting,cross2020macroeconomic,follett2019achieving}. Due to its special connection to frequentist shrinkage priors \citep{polson2010shrink}, it offers not only good finite sample performance but also favourable asymptotic behaviour compared to competing global priors \citep{bhadra2019lasso}. \citet{chakraborty2020bayesian} in particular show that the fractional posterior mean as a point estimator is rate optimal in the minimax sense using (\ref{prior5}). 

Nevertheless, fitting the horseshoe prior can be challenging when the scale parameters are not strongly identified by the data, which is particularly critical in cases where the likelihood is flat, for example, separable data in logistic regression \citep{piironen2017sparsity}.\footnote{We thank an anonymous reviewer for having facilitated this discussion.} We provide in the appendix a (\ref{robustness}) robustness check based on \citet{piironen2017sparsity} that are able to alleviate any identifiability concerns for the empirical study below.

Posteriors are described in the appendix (\ref{subsec:posteriors}).

% 5. SAVS algorithm
\subsubsection{SAVS Algorithm}
Although the horseshoe prior shrinks noise variables towards zero, the importance of a variable for nowcasts may not be immediately clear from posterior summary statistics of the coefficients, especially when the posterior is multi-modal. To aid interpretability and simultaneously preserve predictive ability, we employ the signal adaptive variable selection (SAVS) algorithm of \cite{ray2018signal} to the posterior coefficients on a draw-by-draw basis. The algorithm uses a useful heuristic, inspired by frequentist lasso estimation, to threshold posterior regression coefficients to zero:
\begin{equation}\label{eq:savs1}
    \phi_j = sign(\hat{\beta}_j)||X_j||^{-2}max(|\hat{\beta}_j|\;||X_j||-\kappa_j,0),
\end{equation}
where $X_j = (x_{j1},\cdots,x_{jT})'$ is the $j^{th}$ column of the regressor matrix X, $sign(x)$ returns the sign of $x$ and $\hat{\beta}$ represents a draw from the regression posterior. The parameter $\kappa_j$ in (\ref{eq:savs1}) acts as a threshold for each coefficient akin to the penalty parameter in lasso regression which can be selected via cross-validation. \citet{ray2018signal} propose
\begin{equation}\label{eq:savs2}
    \kappa_j = \frac{1}{|\beta_j|^2},
\end{equation}
which ranks the coefficients inverse-squared proportionally and provides good performance compared to alternate penalty levels \citep{ray2018signal,huber2019inducing}. To see the similarity to lasso style regularisation, the solution to (\ref{eq:savs1}) can be obtained by the following minimisation problem which is closely related to the adaptive lasso \citep{zou2006adaptive}: 
\begin{equation}\label{eq:savs3}
    \overline{\phi} = \underset{\phi}{\mathrm{argmin}}\bigg\{\frac{1}{2}||X\hat{\beta}-X\phi|| + \sum^K_{j=1}\kappa_j|\phi_j|\bigg\}.
\end{equation}
Here, $\overline{\phi}$ is the sparsified regression vector. Analogous to the SSVS posterior, the relative frequency of non-zero entries in the posterior coefficient vector can be interpreted as posterior inclusion probabilities. Integrating over the uncertainty of the parameters, we obtain the predictive distribution $p(\tilde{y}|y)$, which is similar to a Bayesian Model Averaged (BMA) posterior \citep{huber2019inducing}.

% 5. Sampling Algorithm
\subsubsection{Sampling Algorithm}
With the conditional posteriors for the regression and state components at hand (see Appendix \ref{subsec:posteriors}), we sample states as well as regression parameters with the following Gibbs sampler:
\begin{enumerate} \label{gibbssampler}
    \item Sample $(\tilde{\tau},\tilde{\alpha}|y, \theta, \beta,\sigma^2_y)$
    \item Sample $(\theta|y,\beta,\tilde{\tau},\tilde{\alpha},\sigma^2_y)$
    \item Randomly permute signs of $(\tilde{\tau},\tilde{\alpha})$ and $(\sigma_{\tau},\sigma_{\alpha})$
    \item Sample $(\beta|y,\theta,\tilde{\tau},\tilde{\alpha},\sigma^2_y)$
    \item Sample $(\sigma^2_y|y,\tilde{\tau},\tilde{\alpha},\sigma^2_y)$
\end{enumerate}

As mentioned in Section \ref{sub:Non_centred}, states are sampled in a non-recursive fashion which exploits sparse matrix computation and precision sampling. The exact sampling algorithm is given in Appendix  \ref{subsec:estimation}. As discussed in Section \ref{sub:Non_centred}, after sampling $\theta$ in step 2, we randomly permute signs of $(\tilde{\tau},\tilde{\alpha}),(\sigma_{\tau},\sigma_{\alpha})$ to aid mixing. Step 4 of the sampler will depend on the prior and its respective hyperpriors. While the posterior sampling scheme for the SSVS is standard, we use the efficient posterior sampler of \citet{bhattacharya2016fast} to sample the regression coefficients of the horseshoe prior. Compared to Cholesky based sampling as used for the SSVS, computation speed is markedly improved; see Appendix  \ref{subsubsec:hs}. Note that in step 4, we perform SAVS sparsification via (\ref{eq:savs1}) on an iteration basis. 

\section{Data} \label{sec:data}
\subsection{Mixing Frequencies}
% Mixed Frequency
In this paper, we relate monthly macro data commonly used for nowcasting based on \citet{giannone2016exploiting} and internet search information via U-MIDAS skip-sampling to real quarterly U.S. GDP growth. The U-MIDAS approach to mixed frequency belongs to the broader class of `partial system' models \citep{banbura2013now}, which directly relate higher frequency information to the lower frequency target variable by vertically realigning the covariate vector. The benefit of this mixed frequency method compared to restricted MIDAS and full system state space methods is its simplicity in that existing models and priors can directly be applied to U-MIDAS sampled data as well as its competitive performance, especially when the frequency mismatch between the target and the regressors is small \citep{foroni2015unrestricted,foroni2014comparison}, as is the case in our application. Switching notation from equation (\ref{equ:bsts}) to make it explicit that $y_t$ is quarterly while $x_t$ is sampled at a higher, i.e., monthly frequency, denote $x_{t,M}=(x_{1,t,M}, \cdots, x_{K,t,M})$ and $\beta_m = (\beta_{1,M},\cdots,\beta_{K,M})'$ where $M=(1,2,3)$ denotes the monthly observation of the covariate within quarter, $t$. By concatenating each monthly column, we obtain a $T\times 3K$ regressor matrix $\boldsymbol{X}$ and a $3K \times 1$ regression coefficient vector $\boldsymbol{\beta}$. This vertical realignment is visualised for a single representative regressor below:
\begin{equation}  \label{eq:skip-sampling}
  \left( \begin{array}{ccccc}
y_{1st quarter} & | & x_{Mar} & x_{Feb} & x_{Jan} \\
y_{2nd quarter} & | & x_{Jun} & x_{May} & x_{Apr} \\
. & | & . & . & . \\ 
. & | & . & . & . \\
. & | & . & . & . \\ \end{array} \right)  
\end{equation} 
\subsection{Macroeconomic Data}
% Macro Data & Real-Time vintages
The macro data set pertains to an updated version of the database of \citet{giannone2016exploiting} (henceforth, `macro data') which contains 13 time series which are closely watched by professional and institutional forecasters including real indicators (industrial production, house starts, total construction expenditure etc.), price data (CPI, PPI, PCE inflation), financial market data (BAA-AAA spread) and credit, labour and economic uncertainty measures (volume of commercial loans, civilian unemployment, economic uncertainty index etc.). We augment this data set with the composite Purchasing Managers Index (PMI) and the University of Michigan Consumer Confidence Index (UMCI). These are often used as leading indicators for producer and consumer sentiment, respectively. The target variable for this application is deseasonalised U.S. real GDP growth (GDP growth) data as reported in the FRED dataset.\footnote{Here, the deseasonalisation pertains to the X13-ARIMA method and was performed prior to download from the FRED-MD website.}$^{,}$\footnote{We thank an anonymous reviewer who brought to our attention that instead of mixing pre-deseasonalisation techniques between macroeconomic data and Google Trends discussed below, one could also deseasonalise with common techniques such as the Loess filter. In doing so, the results in this paper remain qualitatively identical. Details are available upon request.}

% Real-time Vintages
As early data vintages of macroeconomic data and GDP figures can exhibit substantial variation compared to final vintages \citep{croushore2006forecasting,sims2002role}, there is no unambiguous choice of variable in evaluating nowcast models on historical data. Further complications can arise through changing definitions or methods of measurements \citep{carriero2015realtime}. In order to judge the expected performance of the proposed models from a real-time perspective, we only make use of the latest vintages of the series available at the point in time of the nowcast. \footnote{The only exception is real GDP growth, for which, following previous nowcast studies \citep{carriero2015realtime,clark2011real}, we use the second vintage for nowcast evalutation.} The stylised release calendar (\ref{tab:calendar}) simulates the data availability within the months during which nowcasts are conducted. For instance, at the 24th nowcast period, all data which became available during periods 1-24 will be updated according to their latest available vintages dating prior to the release of PCE and PCEPI, which are published typically during the last week of a given month. Real-time vintages are downloaded from the FRED database using the `FredFetch' Matlab package.\footnote{The Matlab package is available from \url{https://github.com/MattCocci/FredFetch}. PMI data were downloaded from \url{quandl.com} using Quandl code `ISM/MANPMI'.}

% Dependent Variable
%The target variable for this application is deseasonalised U.S. real GDP growth (GDP growth) data as reported in the FRED website.\footnote{Here, the deseasonalisation pertains to the X13-ARIMA method and was performed prior to download from the FRED-MD website.} We found that pre-deseasonalised data improved forecast accuracy compared to modeling it in our state space system. This may be due to the small sample size. As Google Trends were only available from 01/01/2004-01/06/2019 at our time of download, the period under investigation pertains to the same period in quarters (60 quarters). We split the data set into a training sample of 45 quarters (2004q2-2015q2) and a forecast sample of 15 quarters (2015q3-2019q1).

% Macro Data Set
%The macro data set pertains to an updated version of the database of \citet{giannone2016exploiting} (henceforth, ‘macro data’) which contains 13 time series which are closely watched by professional and institutional forecasters including real indicators (industrial production, house starts, total construction expenditure etc.), price data (CPI, PPI, PCE inflation), financial market data (BAA-AAA spread) and credit, labour and economic uncertainty measures (volume of commercial loans, civilian unemployment, economic uncertainty index etc.). Table (\ref{tab:calendar}) gives an overview over all data along with FRED codes.

\subsection{Google Trends}
% Google Trends Data Set. Perhaps expand on the fact that broad GDP measures necessitate a broad range of GT
Google Trends (GT) are indices produced by Google on the relative search volume popularity of a given search term, search topic or pre-specified search category, conditional on a given time frame and location. The difference between individual search terms and topics/categories is that the latter measures the search popularity for a basket of search terms which are content-wise related to the specified topic or category. In particular, categories are further split into a 5-level hierarchy of categories which are fixed a priori,\footnote{For an overview of categories and sub-categories, please see \url{https://github.com/pat310/google-trends-api/wiki/Google-Trends-Categories}} and topics can be assembled depending on the term one is interested in. For example, the user can specify the topic `Recession', whose related search queries contain, among others, `recession', `downturn', and `economic depression'. Likewise, the category `Welfare \& Unemployment' relates to search queries about `unemployment, `food stamps' and `social security office'. A large literature on using individual Google Trends search terms\footnote{See, for example:  \citet{guzman2011internet,mclaren2011using,askitas2009google,fondeur2013can,carriere2013nowcasting}.} have shown that these data can improve predictions for economic time-series which have a clear connection to the specific search term used, such as using `unemployment benefits' to predict unemployment \citep{smith2016google}. However, this approach has two potential limitations. 

First, using broad search terms to capture general macroeconomic activity bears the risk of capturing spurious search behaviour. For example, the search term `jobs' might contain search volume for `Steve Jobs'. Second, since many search terms will be related to multiple topics, there may be lack of interpretability. To reduce search term ambiguity and interpretability in relationship to real GDP growth, we use Google topics and categories instead of individual search terms, and choose these based on their relationship with various aspects of the economy. As forcefully argued by \citet{woloszko2020tracking} and \citet{fetzer2020coronavirus}, these mostly alleviate spuriously correlated search terms as the user can confine the search purpose. This is benefited by the fact that Google refines this basket of search terms, by taking into account where users click after the search has been conducted \citet{woloszko2020tracking}. Further, categories and topics can be conceptualised as factors based on search terms with the same meaning/purpose. Although the exact basket of search terms corresponding to a topic/category is not a priori accessible to the user, any topic or category with little predictive power will ultimately be shrunk to zero via the shrinkage priors employed in the proposed models. 

Our sample comprises 37 Google Trends which were chosen based on capturing activity in various parts of the economy ranging from crisis/recession, labour market, personal finance, consumption to supply side activities.\footnote{This list was inspired by previous research such as \citet{woloszko2020tracking}.} Our chosen list of topics and categories is as follows:

\begin{itemize}
    \item \textbf{Crisis/Recession}: topic - Economic crisis, topic - Crisis, topic - Recession
    \item \textbf{Labour Market}: topic - Unemployment benefits, topic - jobs, topic - Unemployment, Welfare \& unemployment
    \item \textbf{Bankruptcy}: topic - Bankruptcy, topic - foreclosure
    \item \textbf{Credit, Loans \& Personal Finance}: topic - Investment, topic - Mortgage, topic - Interest rate, Credit \& lending, Investing
    \item \textbf{Consumption Items \& Services}: Food \& drink, Vehicle brands, Home \& garden, Sports, Autos \& vehicles, Grocery \& food retailers, Vehicle licensing \& registration, Hotels \& accommodations
    \item \textbf{News}: Business news, Economy news
    \item \textbf{Housing}: topic - Affordable housing, topic - House price index
    \item \textbf{Business \& Industrial Activity}: Construction, consulting \& contracting, Business services, Transportation \& logistics, manufacturing
    \item \textbf{Health}: Health
\end{itemize}
The  relatively large proportion of search items related to consumption of goods and services reflects the large role of consumption in determining U.S. GDP. \citet{vosen2011forecasting} and \citet{woo2019forecasting} have shown that similar search items track and predict the UMCI index and private consumption very well, thereby capturing consumer sentiment. Labour topics and categories track the popularity of search terms related to job search and benefits demand which \cite{smith2016google}, \cite{d2017predictive} and \cite{fondeur2013can} have shown to predict the unemployment rate in various countries. Topics related to personal finance and investment may signal wealth effects \citep{woloszko2020tracking}, which tend to positively correlate with the business cycle (see figure \ref{fig:gtplots} in appendix). Topics around housing have been shown to be indicative of housing prices \citep{wu2015future,askitas2009google}. The recession, business news and bankruptcy themed search items typically increase during economic downturns which therefore act as signals of economic distress and recessions \citep{castelnuovo2017google,chen2012forecasting}.

% Root Term methodology
While the selection of our search items is subjective, in general, there is no consensus on how to optimally select search terms for final estimation. Methods proposed in the previous literature can be summarised as: (i)~pre-screening through correlation with the target variable as found via Google Correlate \citep{scott2014predicting,niesert2020can,choi2012predicting};\footnote{Unfortunately, Google Correlate has suspended updating their databases past 2017.} (ii)~cross-validation \citep{ferrara2019google}; (iii)~use of prior economic intuition where search terms are selected through backward induction \citep{smith2016google,ettredge2005using,askitas2009google}; and (iv)~root terms, which similarly specify a list of search terms through backward induction, but additionally download ``suggested'' search terms from the Google interface. This serves to broaden the semantic variety of search terms in a semi-automatic way. As methodologies based on pure correlation do not preclude spurious relationships \citep{scott2014predicting,niesert2020can,ferrara2019google}, we opt for our (somewhat subjective) selection to best guarantee economically relevant Google Trends. 

% Treatment of Google Trends
Since search terms can display seasonality, we deseasonalise all Google Trends by the Loess filter, as recommended by \citet{scott2014predicting}, which is implemented with the ``stl'' command in R.\footnote{To mitigate against inaccuracy stemming from sampling error, we downloaded the set of Google Trends seven times between 1 August 2021 to 8 August 2021 and took the cross-sectional average. Since we used the same IP address and google-mail account, there might still be some unaccounted measurement errors. However, using topics and categories instead of individual search terms, we observe much lower sampling variance.}

% Pseudo real time calendar
Although one of the main benefits of Google Trends is their timely availability, which can be as granular as displaying search popularity minutely for the past hour, the purpose of the empirical application is to showcase the flexibility of the proposed models in taking advantage of the heterogeneous information contained in adding new data sources to traditional macroeconomic data, even with little data processing efforts. Due to the simplicity of obtaining monthly Google Trends information, we sample the Google Trends information at the monthly frequency. Nevertheless, the proposed methodology can easily be extended to update monthly Google Trends with higher frequency search information via bridge methods as in \citet{ferrara2019google} or could directly be included in the model via expansion of the covariate matrix.\footnote{Due to the already very high-dimensionality of the data set, we retain such extensions for future investigation. Constraining the parameter space via MIDAS sampling might make estimation more feasible.}

The indicative real-time calendar can be found in Table 1 and has been constructed after the data's real publication schedule. It comprises a total of 37 nowcast periods which make for an equal number of information sets $\Omega^v_t$ for $v = 1,\cdots,37$ which are used to construct nowcasts as explained in Section \ref{sec:nowcasting}. Google Trends are treated as released prior to any other macro information pertaining to a given month, since as argued, Google Trends information can essentially be continuously sampled.
\begin{table}
    \centering
    \adjustbox{max height=\dimexpr\textheight-5.5cm\relax,
           max width=\textwidth}{
    \begin{tabular}{l l l l l l l}
    \hline
    Releases & Timing & Release & Variable Name & Pub. lag & Transformation & FRED Code  \\
    \hline
    \hline
    1 & First day of month 1 & No information available & - & - & - & -  \\
    2 & Last day of month 1 & Google Trends & & m & 4 & - \\
    3 & 4th Friday month 1 &  Consumer Sentiment & cons & m & 3 & UMCSENT \\ 
    4 & Last day of month 1 & Fed. funds rate \& credit spread & fedfunds \& baa & m & 3 & FEDFUNDS \& BAAY10 \\
    5 & 1st bus. day of month 2 & Economic Policy Uncertainty Index & uncertainty & m-1 & 1 & USEPUINDXM \\
    6 & 1st bus. day of month 2 & PMI & pmi & m-1 & 1 & - \\ 
    7 & 1st Friday of month 2 & Employment situation & hours \& unrate & m-1 & 2 & AWHNONAG \& UNRATE  \\
    8 & Middle of month 2 & CPI & cpi & m-1 & 2 & CPI \\
    9 & 15th-17th of month 2 & Industrial Production & indpro & m-1 & 2 & INDPRO \\
    10 & 3rd week of month 2 & Credit \& M2 & loans \& m2 & m-1 & 2 & LOANS \& M2 \\
    11 & Later part of  month 2 & Housing starts & housst & m-1 & 1 & HOUST \\
    12 & Last week of month 2 & PCE \& PCEPI & pce \& pce2 & m-1 & 2 & PCE \& PCEPI  \\
    13 & Last day of month 2 & Google Trends & & m & 4 & - \\
    14 & 4th Friday month 2 & Consumer Sentiment & cons & m & 3 & UMCSENT \\ 
    15 & Last day of month 2 & Fed. funds rate \& credit spread & fedfunds \& baa & m & 3 & FEDFUNDS \& BAAY10 \\
    16 & 1st bus. day of month 3 & Economic Policy Uncertainty Index & uncertainty & m-1 & 1 & USEPUINDXM \\
    17 & 1st bus. day of month 3 & PMI & pmi & m-1 & 1 & - \\ 
    18 & 1st bus. day of month 3 & Construction starts & construction & m-2 & 1 & TTLCONS \\
    19 & 1st Friday of month 3 & Employment situation & hours \& unrate & m-1 & 2 & AWHNONAG \& UNRATE  \\
    20 & Middle of month 3 & CPI & cpi & m-1 & 2 & CPI \\
    21 & 15th-17th of month 3 & Industrial Production & indpro & m-1 & 2 & INDPRO \\
    22 & 3rd week of month 3 & Credit \& M2 & loans \& m2 & m-1 & 2 & LOANS \& M2 \\
    23 & Later part of  month 3 & Housing starts & housst & m-1 & 1 & HOUST \\
    24 & Last week of month 3 & PCE \& PCEPI & pce \& pce2 & m-1 & 2 & PCE \& PCEPI  \\
    25 & Last day of month 3 & Google Trends & & m & 4 & - \\
    26 & 4th Friday month 3 & Consumer Sentiment & cons & m & 3 & UMCSENT \\ 
    27 & Last day of month 3 & Fed. funds rate \& credit spread & fedfunds \& baa & m & 3 & FEDFUNDS \& BAAY10 \\
    28 & 1st bus. day of month 4 & Economic Policy Uncertainty Index & uncertainty & m-1 & 1 & USEPUINDXM \\
    29 & 1st bus. day of month 4 & PMI & pmi & m-1 & 1 & - \\ 
    30 & 1st bus. day of month 4 & Construction starts & construction & m-2 & 1 & TTLCONS \\
    31 & 1st Friday of month 4 & Employment situation & hours \& unrate & m-1 & 2 & AWHNONAG \& UNRATE  \\
    32 & Middle of month 4 & CPI & cpi & m-1 & 2 & CPI \\
    33 & 15th-17th of month 4 & Industrial Production & indpro & m-1 & 2 & INDPRO \\
    34 & 3rd week of month 4 & Credit \& M2 & loans \& m2 & m-1 & 2 & LOANS \& M2 \\
    35 & Later part of  month 4 & Housing starts & housst & m-1 & 1 & HOUST \\
    36 & Last week of month 4 & PCE \& PCEPI & pce \& pce2 & m-1 & 2 & PCE \& PCEPI  \\
    37 & Later part of month 5 & Housing starts & housst & m-2 & 1 & HOUST \\
    \hline
    % Note: I cut short after the last variable has been published, as no new data would be published that would explain reference period growth. Hence, I cut the calendar short after PCEIP release.
    \end{tabular}
    }
    \caption{Real time calendar based on actual publication dates. Transformation: 1 = monthly change, 2 = monthly growth rate, 3 = no change, 4 = LOESS decomposition. Pub. lag: m = refers to data for the given month within the reference period, m-1 = refers to data with a months' lag to publication in the reference period, m-2 = refers to data with 2 months' lag to publication in the reference period.}
    \label{tab:calendar}
\end{table}

\subsection{Understanding Google Trends}
\begin{figure}[h]
    \centering
    \includegraphics[width=\textwidth]{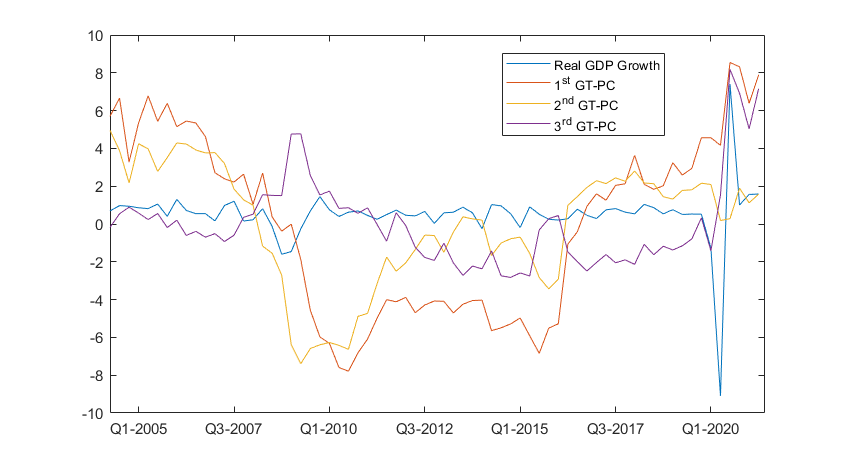}
      \caption{First 3 principal components of the GT data set and 2nd vintage real GDP growth.}
    \label{fig:gt_pca}
\end{figure}

\begin{figure}[h]
    \centering
    \includegraphics[width=\textwidth]{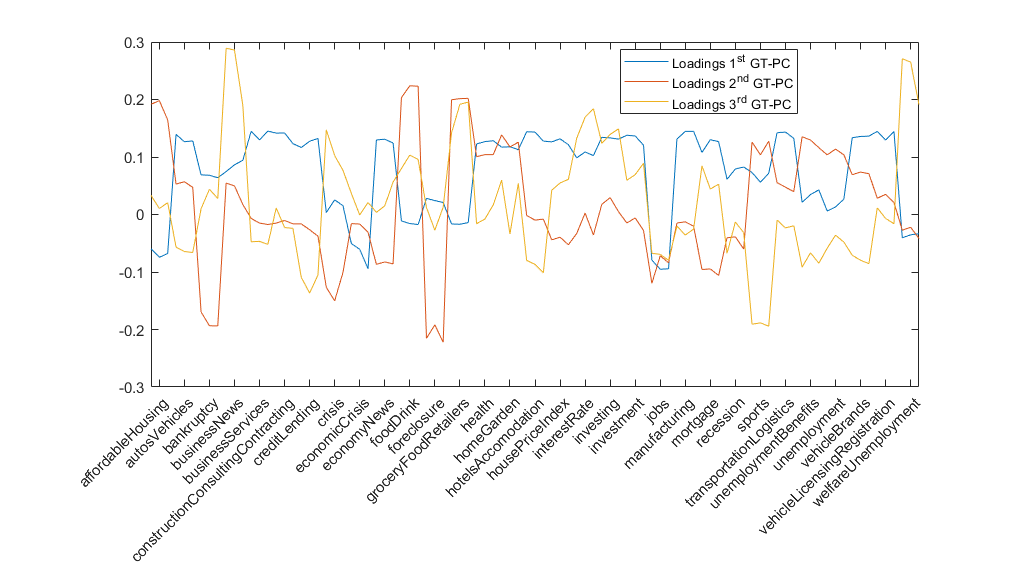}
      \caption{ Loadings for first 3 principal components of the GT data set.}
    \label{fig:gt_load}
\end{figure}

To get a visual understanding of the Google search information, we plot in figure (\ref{fig:gt_pca}) the first 3 principal components of the U-MIDAS transformed Google Trends information. Figure (\ref{fig:gt_load}) shows the factor loadings.

The first three principal components show very heterogeneous behaviour, but the dynamics conform to the economic intuition suggested from the loadings. The first component loads positively on supply side activity such as `Business services', `Construction, consulting \& contracting', `Manufacturing' as well as `Investing' and negatively on recessionary themes and `Jobs' which spike during the crises and decrease during recoveries (see figure \ref{fig:gtplots} for indicative time-series plots of individual Google search items). Accordingly, the first component decreases strongly during the financial crisis (and to a smaller degree also during height of the Covid-19 recession) and picks up the rapid increase in economic activity after Q2 2020 very well.

The second component can be understood as a measure of consumer sentiment and financial health as it loads mostly on consumption items and negatively on topics such as `Bankruptcy' and `Foreclosure'; both terms were very popular during the financial crisis, but not so much the pandemic crisis.\footnote{This may reflect the different nature of the Covid-19 pandemic induced recession and the positive impact government policies had.} Accordingly, the second principal component shows a large dip during the financial but only a minor dip during the Covid-19 recession.

The third principal component loads very strongly on business news/recession/crisis items which increase in popularity during periods characterised by economic anxiety, thus spiking around the financial crisis and the pandemic period. Hence, it can be interpreted as an indicator of economic distress.

\subsection{Relationship Between Macroeconomic and Google Search Data}
\begin{figure}[h]
    \centering
    \includegraphics[width=\textwidth]{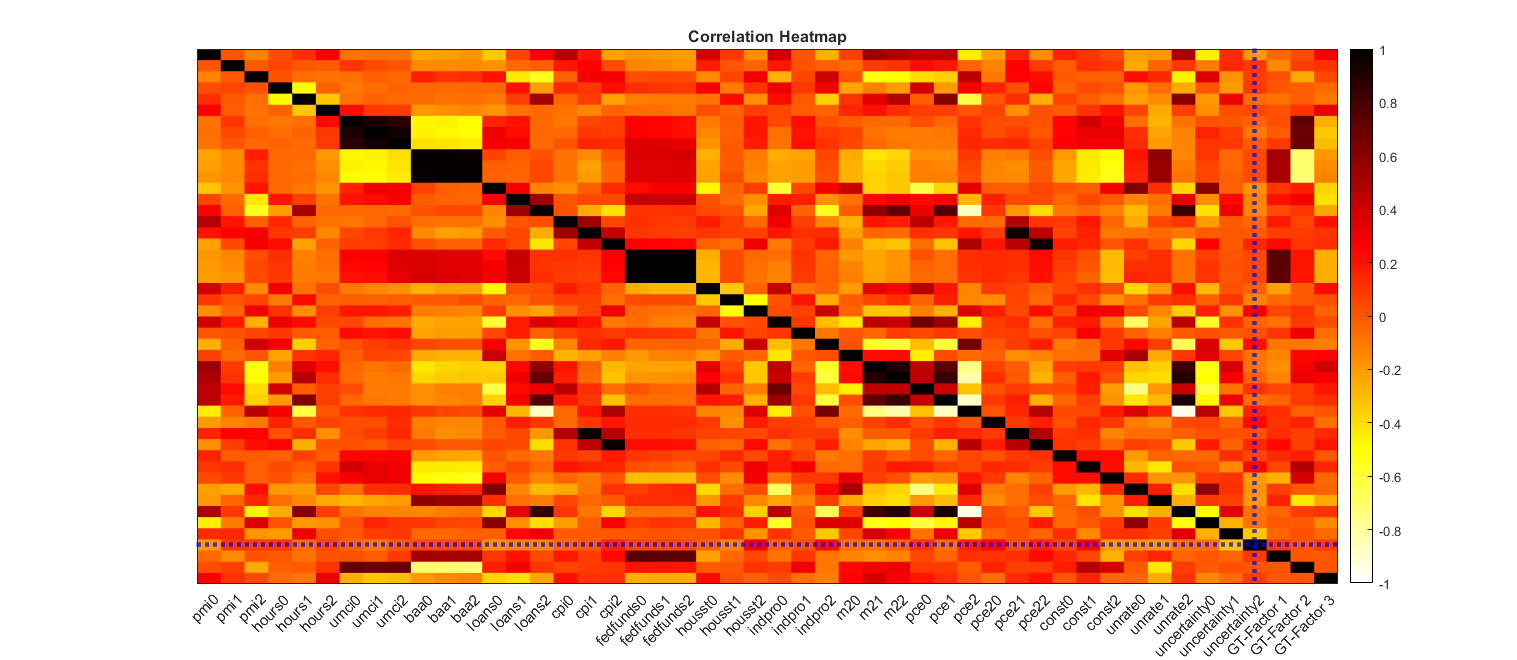}
      \caption{Heatmap of sample correlation matrix.}
    \label{fig:corr_heat}
\end{figure}

Understanding further how the skip-sampled Google Trend series correlate with the macro data set may help us anticipate which information will likely be picked up in the model. Figure (\ref{fig:corr_heat}) shows a correlation heatmap between the macro data set and the first three Google Trends principal components. Please note that an increase in the UMCI and PMI indicate improved consumer and producer sentiments, respectively. 

As expected, the first rather procyclical component correlates highly with the fed-funds rate which tends to also rise with the business cycle. The second component is strongly positively correlated with the UMCI index and negatively with the BAA spread (which increases during deteriorating financial conditions), indicating that it indeed captures something close to consumer sentiment and financial health. Since the third component acts as a recession signal, which abruptly spikes during crises, but is otherwise flat, it is not surprising that macroeconomic variables do not correlate very highly with it. This also indicates that search items in this group might add information that is not well captured by the other included macroeconomic information.

\section{Nowcasting U.S. Real GDP Growth}\label{sec:nowcasting}

The predictive model used to generate in-sample and out-of-sample predictions is given in equations (\ref{eq:non-centred1})-(\ref{eq:non-centred2})
where the first $T=45$ observations are used as training sample.\footnote{As a further alternative to the proposed BSTS models, we investigate in the appendix as well whether past GDP growth dynamics are more appropriately (in terms of nowcasting) modelled via ARMA components. The results show that the LLT components within the BSTS model are clearly preferred over ARMA type dynamics. We thank an anonymous reviewer for making this suggestion.} We estimate three variants of the model based on priors (\ref{eq:nigprior}, \ref{prior5}, and \ref{eq:savs1}) and the original BSTS model of \citet{scott2014predicting}, as well as an AR(4) model for comparison. In line with standard BSTS applications, we first compare the in-sample cumulative absolute one-step-ahead forecast errors, generated from the state space, as well as inclusion probabilities of the variables so as to shed light on which variables produce better fit and explain the outcome. Out-of-sample nowcasts are generated from the posterior predictive distribution $p(y_{T+1}|\Omega^v_{T})$ for growth observation $y_{T+1}$, conditional on the real-time information set $\Omega^v_{T}$, where $(v=1,\cdots,37)$ refers to nowcast periods within the real-time calendar (Table \ref{tab:calendar}). This results in 37 different nowcasts which are generated on a rolling basis until the end of the forecast sample, $T_{end}$. As recommended by \citet{carriero2015realtime}, variables that have not yet been published until nowcast period $v$ are zeroed out. 

Point forecasts are computed as the mean of the posterior predictive distribution and are compared via real time root-mean-squared-forecast-error (RT-RMSFE) which are calculated for each nowcast period as:
\begin{equation} \label{eq:evaluation1}
    \text{RT-RMSFE} = \sqrt{\frac{1}{T_{end}}\sum^{T_{end}}_{j=1}(y_{T+j}-\hat{y}^v_{T+j|\Omega^v_{T+j-1}})^2},
\end{equation}
where $\hat{y}^v_{T+j|\Omega^v_{T+j-1}}$ is the mean of the posterior prediction for nowcast period $v$ using information until $T+j-1$. \\

Forecast density fit is measured by the mean real-time log-predictive density score (RT-LPDS) and real-time continuous rank probability score (RT-CRPS):
\begin{equation} \label{eq:evaluation2}
    \begin{split}
        \text{RT-LPDS} & = \frac{1}{T_{end}}\sum^{T_{end}}_{j=1}logp(y_{T+j}|\Omega^v_{T+j-1}) \\
        & = \frac{1}{T_{end}}\sum^{T_{end}}_{j=1}log \int p(y_{T+j}|\Omega^v_{T+j-1},\zeta_{1:T+j-1})p(\zeta_{1:T+j-1}|\Omega^v_{T+j-1})d\zeta_{1:T+j-1} \\
        & \approx \frac{1}{T_{end}}\sum^{T_{end}}_{j=1}log \left(\frac{1}{M}\sum^M_{m=1}p(y_{T+j}|\Omega^v_{T+j-1},\zeta^m_{1:T+j-1})\right),
    \end{split}
\end{equation}
\begin{equation} \label{eq:evaluation3}
    \text{RT-LPDS} = \frac{1}{T_{end}}\sum_{j=1}^{T_{end}}\frac{1}{2} \left| y_{T+j}-y^{v}_{T+j|\Omega^{\nu}_{T+j-1}} \right| - \frac{1}{2} \left| y^{v,A}_{T+j|\Omega^{v}_{T+j-1}}-y^{v,B}_{T+j|\Omega^{v}_{T+j-1}} \right|,
\end{equation}
where, for brevity of notation, $\zeta_{1:T+j-1}$ collects all model parameters as defined for each model, which are estimated with expanding in-sample information until $T+j-1$ and M stands for iterations of the Gibbs sampler after burn-in. Note that in (\ref{eq:evaluation3}), $y^{v,A,B}_{T+j|\Omega^{v}_{T+j-1}}$ are independently drawn from the posterior predictive density $p(y^{v}_{T+1|\Omega^{v}_{T+j-1}}|y_T)$.

% 2. Pseudo One-step-ahead predictive densities (Might need to be relegated to the appendix)
As shown by \citet{fruhwirth1995bayesian}, in a setting where time-varying and fixed components for a structural state space model are chosen, the LPDS can be interpreted as a log-marginal likelihood based on the in-sample information and therefore provides a model founded scoring rule. The RT-CRPS can be thought of as the probabilistic generalisation of the mean-absolute-forecast-error. Similar to the log-score, it belongs to the broader class of strictly proper scoring rules \citep{gneiting2007strictly} which allows for comparing density forecasts in a consistent manner.\footnote{We do not report calibration tests, as there are too few out-of-sample observations to meaningfully determine calibration.}$^,$\footnote{Although the CRPS is a symmetric scoring rule, it penalises outliers less aggressively than the log-score which is of advantage in small forecast samples such as ours.} To facilitate discussion, our objective is to maximise the RT-LPDS and minimise the RT-CRPS. For all forecast metrics, the predictive distribution used for (\ref{eq:evaluation1}, \ref{eq:evaluation2}, and \ref{eq:evaluation3}) is traditionally generated in state space models via the prediction equations of the Kalman filter \citep{harvey1990forecasting}. Instead, we use the simpler approximate method of \citet{cogley2005bayesian}, which we found to make no practical difference in our sample.\footnote{The Kalman filter provides conditionally optimal forecast densities in terms of squared forecast error. However, if there is mispecification or if the forecast horizon is very short, then approximate methods can do just as well empirically. A similar logic holds when comparing direct and iterative forecasting methods such as in \citet{marcellino2006comparison}. We thank an anonymous reviewer for bringing this to our attention.} The method is described in Appendix \ref{forecasting}. 

Finally, to test whether a state variance is equal to zero, we make use of the Dickey-Savage density ratio evaluated at $\sigma^{\tau,\alpha}$ = 0:
\begin{equation} \label{eq:DS}
    \text{DS} = \frac{p(\sigma^{\tau,\alpha}=0)}{p(\sigma^{\tau,\alpha}=0|y)}
\end{equation}
It can be shown that for nested models, the DS statistic is equivalent to the Bayes factor  between the prior and the posterior distribution of the parameter of interest at zero \citep{verdinelli1995computing}. The intuition for the test is simple: if the prior probability-density-function (PDF) allocates more mass at 0 than the posterior at that point, there is evidence in favour of the unrestricted model, i.e., $\sigma^{\tau,\alpha} \neq 0$. While the priors for the state variances have well known forms and thus can be evaluated analytically, we estimate the denominator for all models through Monte Carlo integration.

% 3. In-Sample Results
\subsection{In-Sample Results}

\begin{figure}[H]
    \centering
    \includegraphics[width=\textwidth]{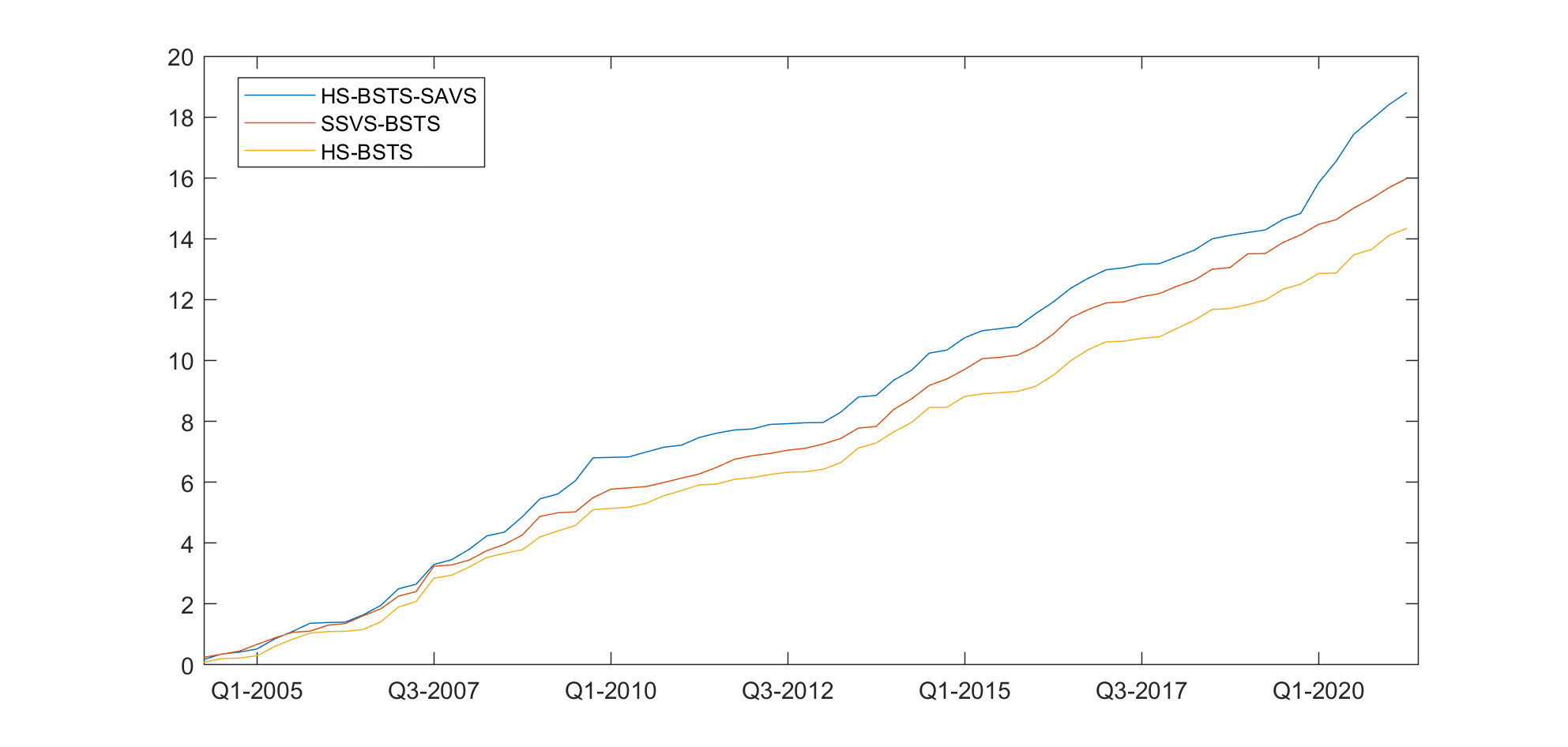}
    \caption{Cumulative one-step-ahead forecast errors in-sample from 3 different models: (1) SSVS-BSTS, (2) HS-BSTS and (3) HS-BSTS-SAVS} 
    \label{fig:cum_errors}
\end{figure}

Figure \ref{fig:cum_errors} shows the in-sample cumulative-one-step ahead prediction errors using the proposed priors where the information set pertains to the entire estimation sample without ragged edges. From Figure \ref{fig:cum_errors}, it is clear that the horseshoe prior BSTS (HS-BSTS) provides the best in-sample predictions at all time periods. The HS-BSTS-SAVS and SSVS-BSTS initially provide similar fit, however diverge in performance around the financial and Covid-19 crises, especially the HS-SAVS-BSTS. It is striking that, compared to the former two, the HS-BSTS provides very stable performance as indicated by a nearly linear increase in errors even during the financial crisis and the Covid pandemic. It is also apparent that the SAVS algorithm is not able to retain the fit of the HS prior alone, which, as we show in the next subsection, is in contrast to the out-of-sample results.  \\

\begin{figure}[H]
    \centering
    \includegraphics[width=\textwidth]{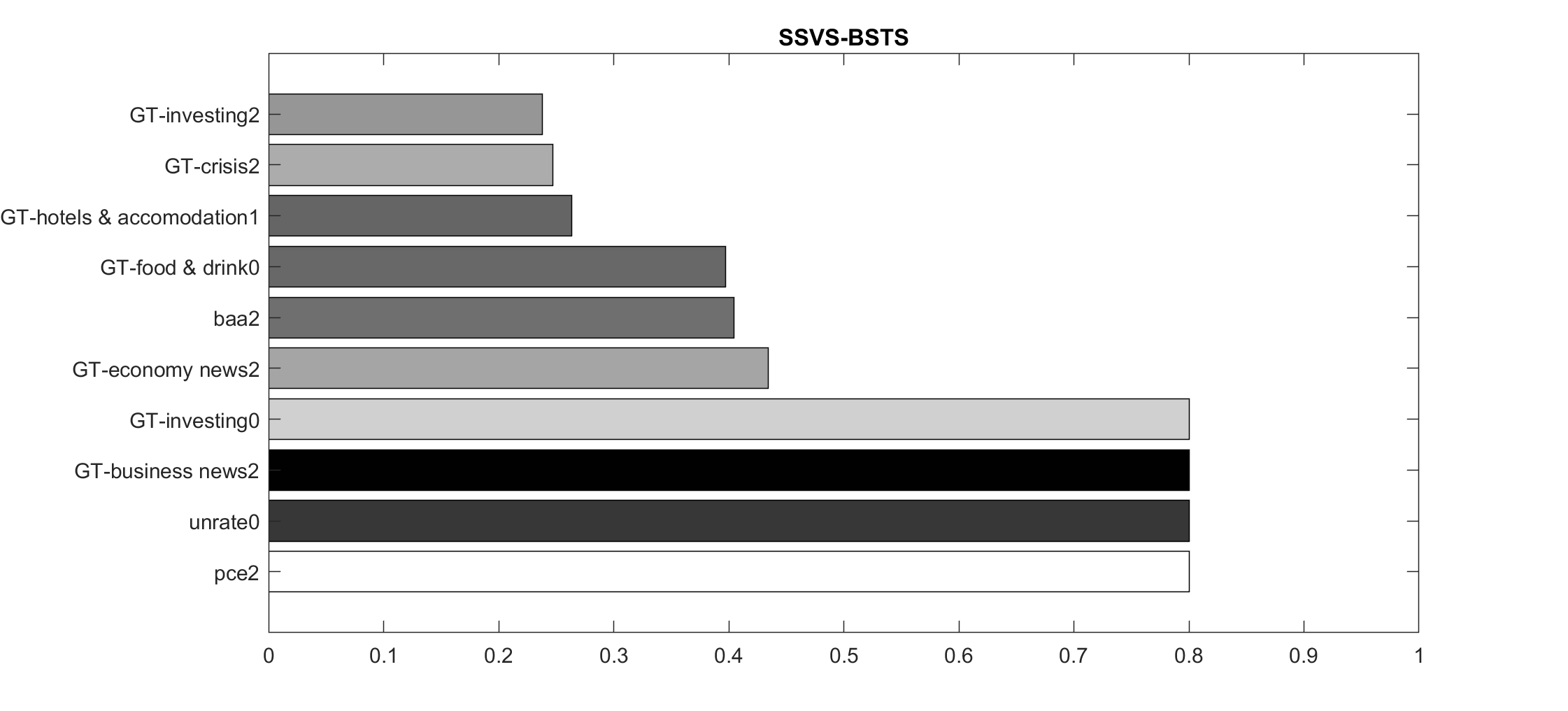}
    \caption{Posterior inclusion probabilities of the SSVS-BSTS model.} 
    \label{fig:PIPSSVS}
\end{figure}

\begin{figure}[H]
    \centering
    \includegraphics[width=0.97\textwidth]{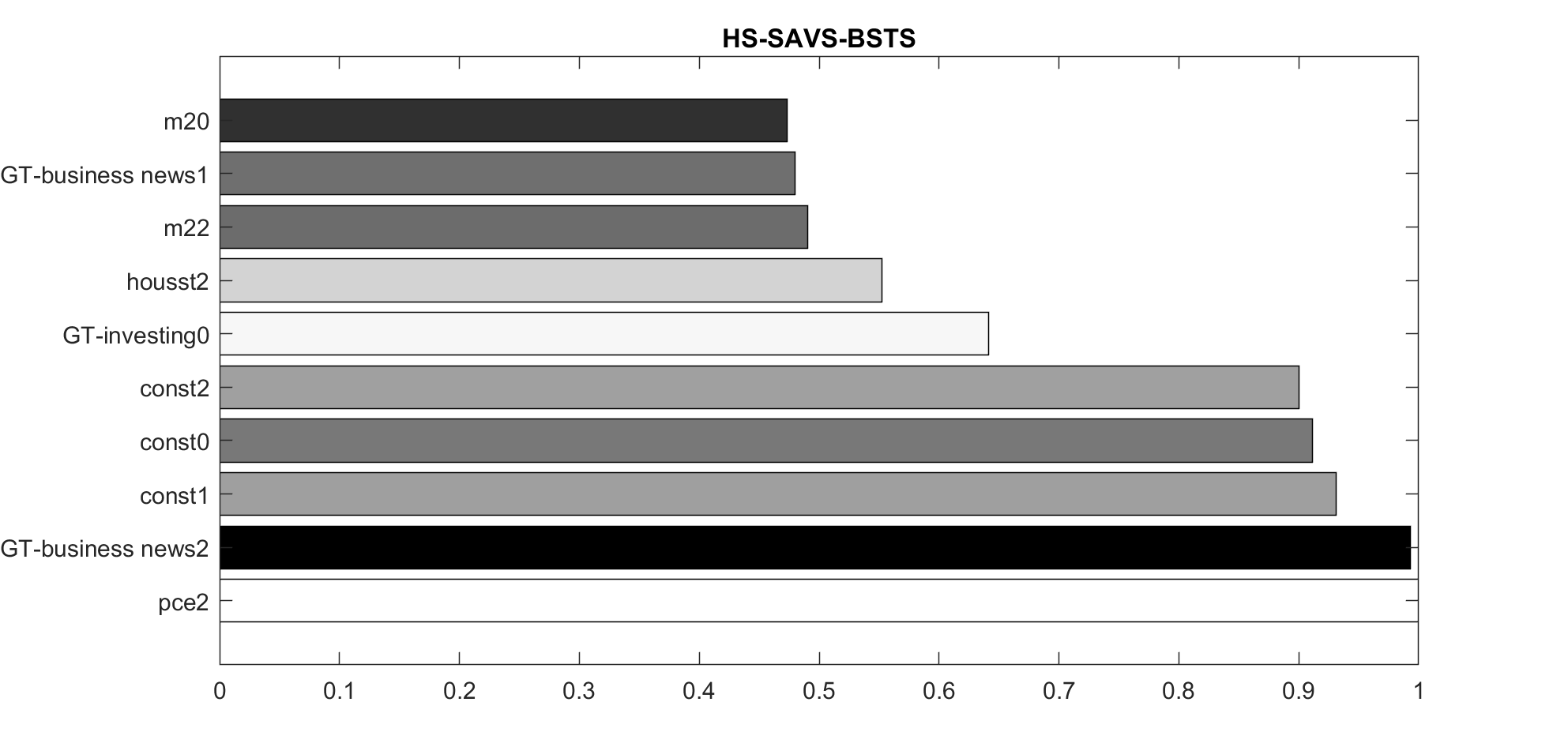}
    \caption{Posterior inclusion probabilities of the HS-SAVS-BSTS model.} 
    \label{fig:PIPHS}
\end{figure}

To understand the driving variables behind the posterior predictive distributions, the posterior marginal inclusion probabilities for the SSVS and SAVS model are plotted for the top ten most drawn variables in Figures \ref{fig:PIPSSVS} and \ref{fig:PIPHS} respectively. The colors of the bars indicate the sign on a continuous scale of white (positive relationship) to black (negative relationship) of the variable when included in the model, and the prefix `GT' indicates Google Trend variables. The number [0,1,2] appended to a variable indicates the temporal position within a given quarter, with 0 being the latest month. 

For all three models (for the BSTS model, see Figure \ref{fig:PIPBSTS}), the posterior inclusion probabilities show that the most drawn Google search information pertains to the category `business news' and topic `investing' with clearly negative and positive impact respectively on GDP growth. The posteriors on these Google Trends variables (Figure \ref{fig:post_gt}) show that not only is the impact statistically significant but also economically so. As search intensity for business news goes up (down), GDP growth forecasts are adjusted downwards (upwards). Vice versa for the investing topic. Since the `business news' category spikes in recessions, but is otherwise flat, this suggests that during periods of heightened recessionary probability and economic fear, people engage and search more for business news which therefore acts as an indicator of expected economic distress. Similar reasoning has led to a large literature on using economic sentiment extracted from news media to model and forecast economic activity \citep{Kalamara2019,kalamara2022making,baker2016measuring,aprigliano2021power,gentzkow2019text,alexopoulos2015power,manela2017news,nyman2021news,shapiro2020measuring}.\footnote{It would be interesting for future research to investigate whether Google Trends and news sentiment extracted from articles substitute or complement each other in modelling recessionary risks.}

Conversely, `investing' items are presumably searched more often when households and individuals are financially more stable which is when they engage in looking for investment opportunities. As seen in Figure \ref{fig:gtplots}, this series tends to positively correlate with the business cycle. This interpretation of the investment topic corroborates findings of \citet{woloszko2020tracking} who show that the investment topic has a positive impact in a panel data nowcasting exercise for GDP growth.

Figures \ref{fig:PIPSSVS} and \ref{fig:PIPHS} also reveal some interesting patterns about how macroeconomic data are employed in the models. The SSVS prior tends to select only the most dominant of the skip-sampled information, while the SAVS extended HS prior allocates significant inclusion probability to all months within a quarter. For example, while the SSVS prior selects the variable `unrate0', i.e., the unemployment rate for the last month in a given quarter, the HS-SAVS prior allocates nearly the same inclusion probability to data for all months on construction starts and M2. This result is likely driven by the fact that the SSVS prior discretises the model space and therefore, with correlated data, will tend to include only the variable with the highest marginal likelihood. Continuous shrinkage priors, on the other hand, make use of all covariates since they are always included in the model.

\begin{figure}[h]
     \centering
 
         \includegraphics[width=\textwidth]{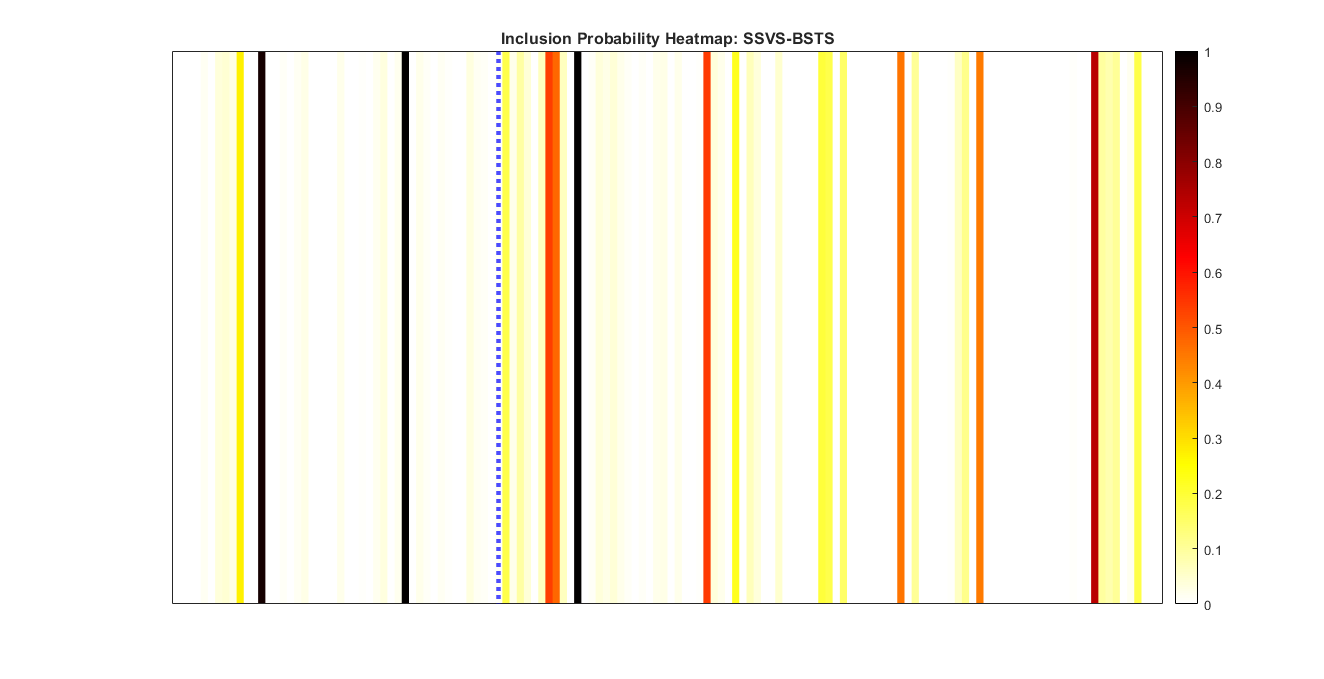}
 
     \caption{Posterior Inclusion probability heatmaps for the SSVS-BSTS. Inclusion probabilities to the left of the dashed line pertain to the macro data set, and to the right, the Google search data.}
     \label{fig:heat_pip_ssvs}
\end{figure}

\begin{figure}[h]
    
         \centering
         \includegraphics[width=\textwidth]{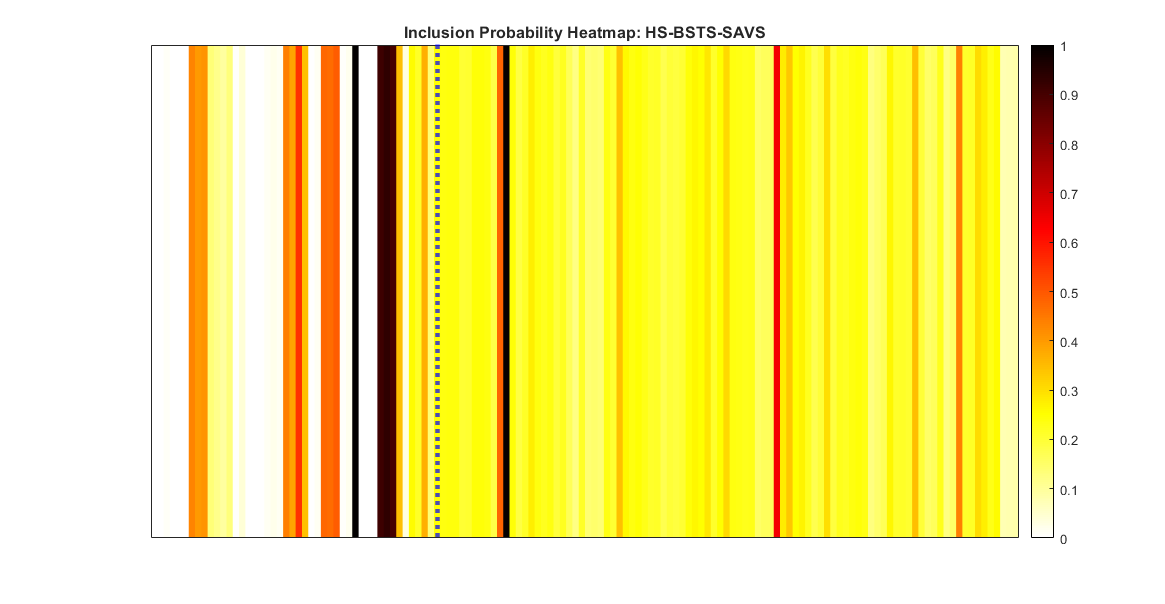}
     \caption{Posterior Inclusion probability heatmaps for the HS-SAVS-BSTS. Inclusion probabilities to the left of the dashed line pertain to the macro data set, and to the right, the Google search data.}
     \label{fig:heat_pip_hs}
\end{figure}

In line with this discussion, the posterior inclusion probability heatmaps in Figures \ref{fig:heat_pip_ssvs} and \ref{fig:heat_pip_hs} show generally that the SSVS and HS prior also display very different degrees of model uncertainty. Note that in the figures, inclusion probabilities to the left of the dashed line pertain to the macroeconomic data set, and to the right, the Google search data. The HS prior tends to display substantial uncertainty over inclusion, particularly for the Google Trends data, which makes sense given the similarity in signal within Google Trends categories and topics. By contrast, the SSVS tends to load only on a few Google search items and not explore posteriors of correlated GTs.

Nonetheless, the fact that the HS prior identifies individual macroeconomic series such as `pce2' as a clear signal indicates that mixed frequency information matter for predictive purposes, which would otherwise be lost in averaging information across quarters.

These results contribute to the recently popularised studies of sparsity within economic prediction problems \citep{giannone2021economic,cross2020macroeconomic} in at least two ways. Firstly, they indicate that different sparsity patterns can emerge within data sources (here, macroeconomic and Google Trends) and within mixed frequency information. And secondly, different sparsity patterns can emerge depending on the prior used. Section \ref{sec:sim} further investigates robustness of the proposed priors to different sparsity settings, and provide recommendations.

\begin{figure}[H]
     \centering
     \begin{subfigure}[b]{0.49\textwidth}
         \centering
         \includegraphics[width=\textwidth]{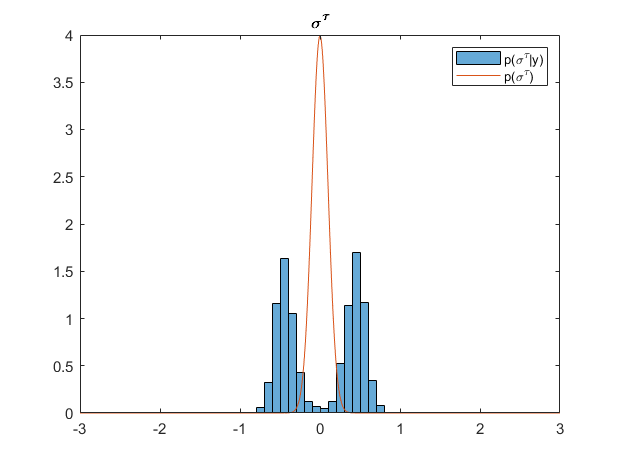}
     \end{subfigure}
     \hfill
     \begin{subfigure}[b]{0.49\textwidth}
         \centering
         \includegraphics[width=\textwidth]{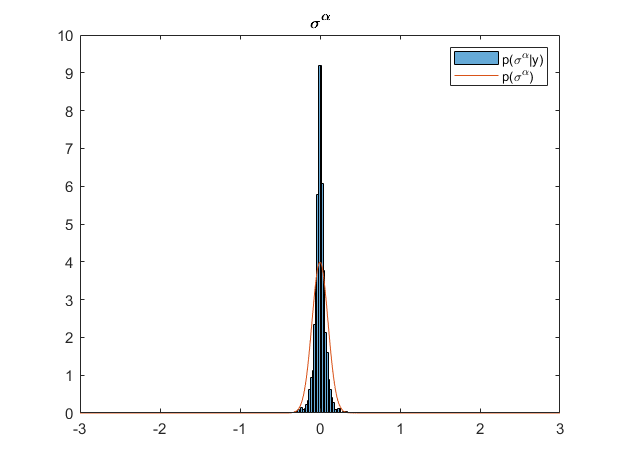}
     \end{subfigure}
     \caption{Distribution of the (left) trend state standard deviation and (right) slope standard deviation for the HS-BSTS model.}
     \label{fig:DS}
\end{figure}

Finally, our in-sample results clearly demonstrate (Figure \ref{fig:DS}) that there is support in the data for a local trend, but not a local linear trend: the posterior for $\sigma^{\tau}$ is clearly bi-modal with less mass on zero than the prior, while the posterior for $\sigma_{\alpha}$ has substantially more mass on zero than the prior. The Bayes factors are 28.69 and 0.42 for the state standard deviations respectively.

%This is confirmed in figure (reference figure 3 which plots histograms of the posterior model sizes). The fact that monthly skip-sampled variables appear to have similar inclusion probabilities shows their explanatory power is similar which therefore leads to model uncertainty. For instance, the construction, BAA as well as Housst variables have similar inclusion probabilities. The SSVS prior instead tends to select only the most dominant of the skip-sampled information. Further, both priors select the Google Trend term ’real GDP growth’ as the most important Google Trend term. As expected, the posterior mean coefficient value of this Google Trend term is negative which gives tentative evidence that this search term acts as a warning signal to downturns of GDP growth as alluded to in the introduction. \\

% 4. Out-of-Sample Results
\subsection{Nowcast Evaluation: Pre-Pandemic period}
We now turn to out-of-sample nowcasting performance, where nowcasts are produced following the real time data publication calendar as explained in Section 2. Due to the extraordinary economic circumstances of the Covid-19 pandemic, we split the evaluation sample into: (a) pre-Covid (ending Q4 2019); and (b) during Covid (ending Q2 2021). We first evaluate point- and then density fit for the pre-Covid period. 

\begin{figure}[H]
    \centering
    \includegraphics[width=\textwidth]{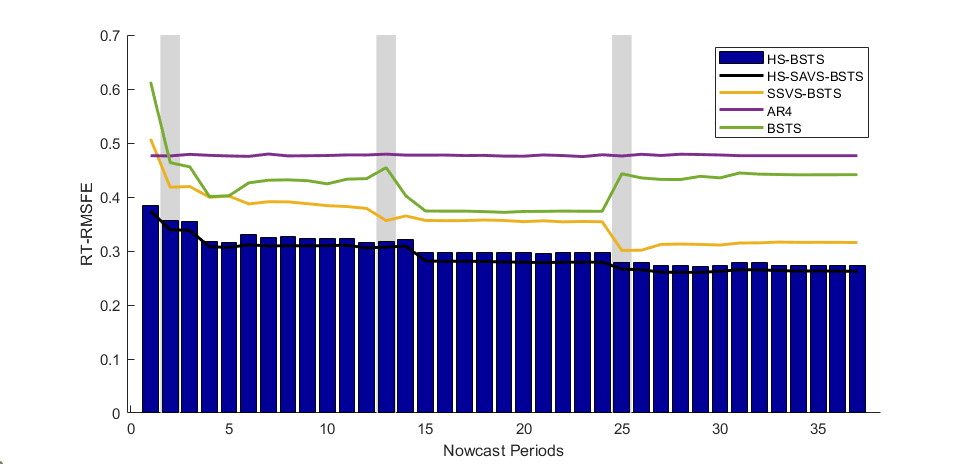}
    \caption{Real-Time RMSFE of all competing models. The RT-RMSFE for the BSTS are plotted on the right axis. Grey vertical bars indicate nowcast periods in which GT are published.}
    \label{fig:RTRMSFE}
\end{figure}

% 4.1: Point-Nowcasts
RT-RMSFE are plotted in Figure \ref{fig:RTRMSFE} for the competing non-centred BSTS estimators, as well as the AR(4) benchmark and the original BSTS model. Note that in all nowcast figures, we represent nowcast periods in which Google Trends are published by grey vertical bars. The following points emerge from Figure \ref{fig:RTRMSFE}. Firstly, it is clear that all proposed BSTS models based on the non-centred state space offer large performance gains (for certain nowcast periods up to 40\%) over the original BSTS model. Secondly, all models nearly monotonically increase in precision as more data are released, where, as expected, the BSTS models outperform the AR benchmark\footnote{The HS-BSTS and HS-SAVS-BSTS outperform the AR model at all nowcast periods significantly as measured by the \citet{diebold1998vevaluating} test at conventional significance levels. The SSVS-BSTS model does so only with the second nowcast period.} as soon as the first data becomes available. Thirdly, among non-centred BSTS models, the HS-SAVS-BSTS does the best; however, it is closely followed by the HS-BSTS. This indicates that the SAVS algorithm successfully shuts down contributions of noise variables and thus gives further validity to the variable selection results discussed above. With only a modest decrease of 2-5\% in RT-MSFE relative to the plain HS-BSTS model, it is evident, however, that the horseshoe prior already provides aggressive shrinkage. Compared to the SSVS-BSTS, the horseshoe prior based BSTS models offer 7-25\% improvements in terms of point forecast accuracy, especially so in the beginning nowcast periods. 

\begin{figure}[H]
    \centering
    \includegraphics[width=\textwidth]{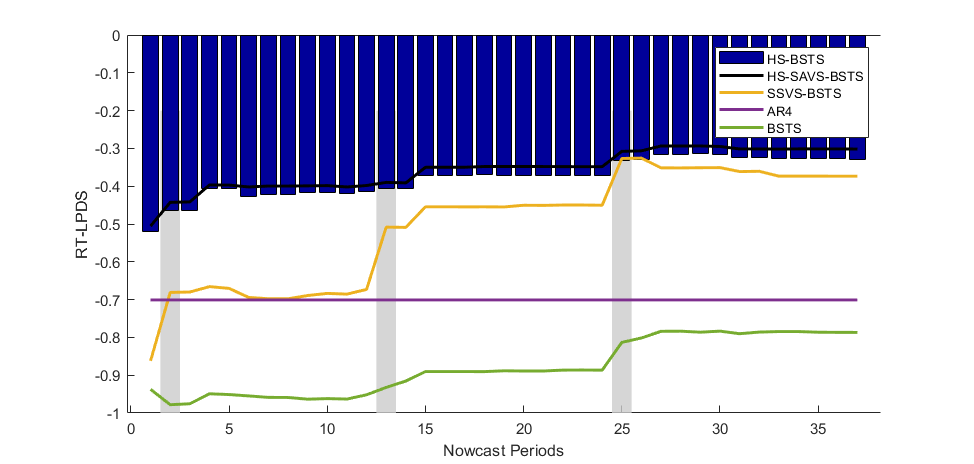}
    \caption{Real-Time log-predictive density scores (RT-LPDS) for all competing models. The RT-LPDS for the BSTS are plotted on the right axis. Grey vertical bars indicate nowcast periods in which GT are published.}
    \label{fig:RTLDPS}
\end{figure}

\begin{figure}[H]
    \centering
    \includegraphics[width=\textwidth]{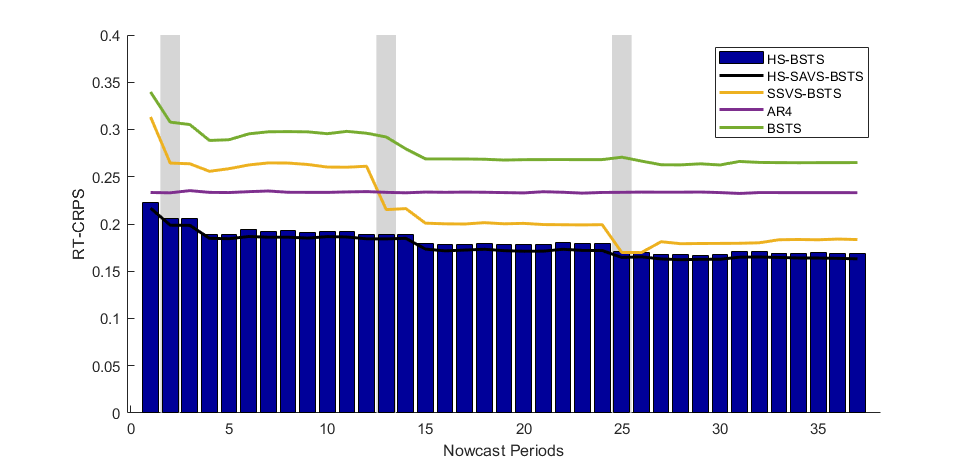}
    \caption{Real-Time CRPS scores (RT-CRPS) for all competing models. The RT-CRPS for the BSTS are plotted on the right axis. Grey vertical bars indicate nowcast periods in which GT are published.}
    \label{fig:RTCRPS}
\end{figure}

Finally, we find that there is a large decrease in point-forecast error due to Google Trends releases prior to macroeconomic data becoming available which is consistent across all models considered. Improvements are in the range of 7-25\% for the given models compared to the first period nowcasts.\footnote{Strictly speaking, these nowcasts are also forecasts due to the information set containing only information from the previous quarter.} The subsequent value of Google Trends for point forecasts is a function of how much a given model loads on the Google Trends search variables and how well the shrinkage prior can separate signal from noise. Hence, improvements for the HS-BSTS models are modest after the first GT release and the SSVS-BSTS experiences a noticeable improvement of 15\% in the final GT release, whereas the original BSTS model of \citet{scott2014predicting} becomes less precise with latter GT releases. The explanation is that the original BSTS model generally struggles with the dimensionality of data set which leads to ineffective variable selection and consequently poor nowcasting performance. \\
%4.2: Density-Nowcasts
Similar to the real-time point forecasts, we plot real-time LPDS (RT-LPDS) and CRPS (RT-CRPS) in Figures \ref{fig:RTLDPS} and \ref{fig:RTCRPS}, respectively. The RT-LPDS and RT-CRPS mostly confirm the main findings from the point nowcasts. In contrast to the point nowcasts, however, there is now a much more clear cut performance improvement in density fit when the Google Trends information are released in periods 15 and 27, especially so for the BSTS model of \citet{scott2014predicting}. This divergence in performance hints at the fact that part of the value of including Google Trends information is to better characterise forecast uncertainty which in turn aids density calibration. Next, we explore this distinctive feature of Google Trends releases for nowcasts during the pandemic.

\subsection{Nowcast Evaluation: During the Pandemic}
The pre-Covid results highlighted that the value of Google Trends are largest before any macroeconomic information are available for the given quarter. While the aim of the nowcasting application for the proposed models is not to provide very granular (weekly or higher) nowcasting models,\footnote{Higher frequencies would expand the covariate set within the U-MIDAS sampling framework even further. At higher frequencies, predictions could instead be based on single covariates and combined, for example via Bayesian model averaging, or alternatively the frequency weights could be constrained via lower parametric basis functions which is akin to conventional MIDAS estimation.}$^{,}$\footnote{See \citet{woloszko2020tracking} for a panel data approach to a weekly GDP index based on Google Trends.} we now illustrate briefly how, even at the relatively coarse monthly level, Google search information improves predictions during the pandemic. 

\begin{figure}[h]
     \centering
     \begin{subfigure}[b]{0.49\textwidth}
         \centering
         \includegraphics[width=\textwidth]{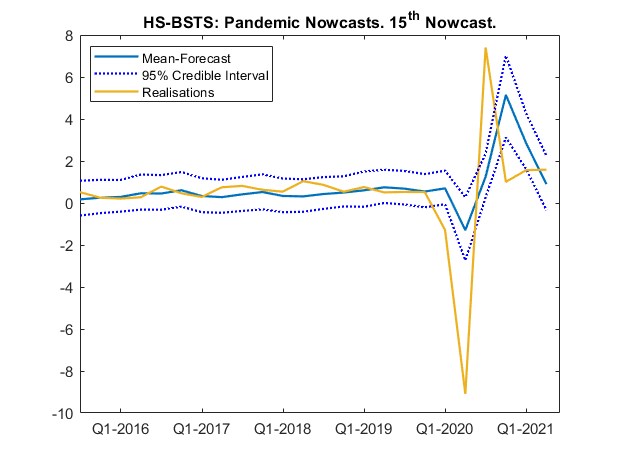}
         \label{fig:y equals x}
     \end{subfigure}
     \hfill
     \begin{subfigure}[b]{0.49\textwidth}
         \centering
         \includegraphics[width=\textwidth]{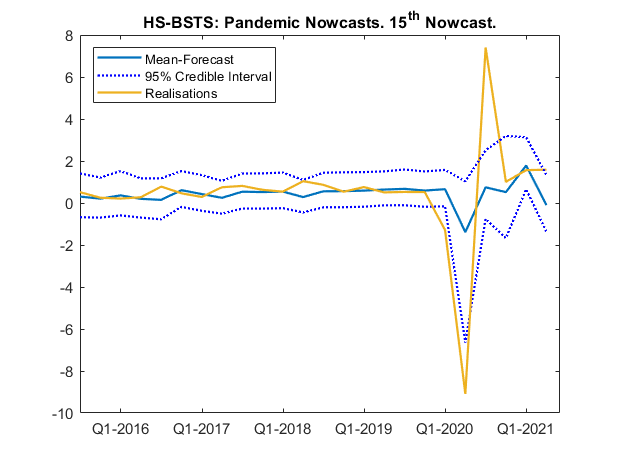}
         \label{fig:three sin x}
     \end{subfigure}
     \hfill
     \begin{subfigure}[b]{0.49\textwidth}
         \centering
         \includegraphics[width=\textwidth]{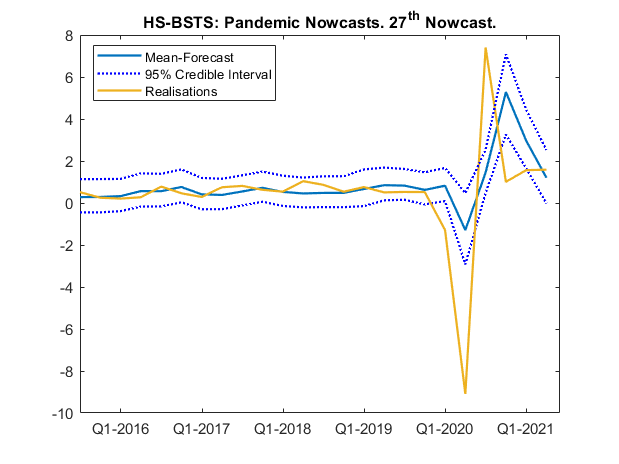}
         \label{fig:five over x}
     \end{subfigure}
     \hfill
         \begin{subfigure}[b]{0.49\textwidth}
         \centering
         \includegraphics[width=\textwidth]{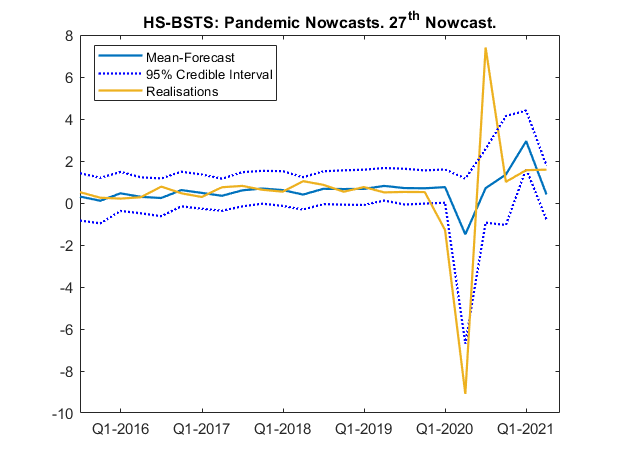}
         \label{fig:five over x}
     \end{subfigure}
        \caption{Predictive distributions for the macro only data set (left column) and the macro + Google Trends data set (right columns) for the 15th  (upper row) and 27th (lower row) nowcast period.}
        \label{fig:pandemic}
\end{figure}

Figure \ref{fig:pandemic} plots mean forecasts with their credible 95\% intervals for the HS-BSTS model based on only macroeconomic data (left column) and the full data set (right column) for the 15th (upper row) and 27th (lower row) nowcast periods respectively. The nowcast periods were chosen to showcase the best possible nowcasts based on information from the end of the second and third month within a quarter respectively. While neither model is able to capture the full extent of the trough during the Covid-19 recession, the model including search term information provides a clear sense of  heightened downside risks through a large asymmetric dip of the lower part of credible forecast interval. This is in line with the findings from \citet{woloszko2020tracking} that for many OECD countries, Google Trends information is able to provide timely downside risk indications. Uniquely, our nowcast exercise highlights how Google Trends can indicate large downward swings in GDP growth over and beyond contributions from macroeconomic data. The fact that Google search information has a greater impact on forecast uncertainty rather than point forecasts further indicates that future research should investigate the potential benefits of using alternative data sources for modelling conditional heteroskedasticity such as in GARCH or stochastic volatility type models.

\subsection{An Extension to Student-t Errors}
It is also clear that both models struggle to nowcast the equally large upswing that follows the pandemic trough. Inspired by recent VAR forecasting literature during the pandemic \citep{lenza2020estimate,carriero2021addressing}, we also explore a new BSTS model based on fatter tailed t-distributed errors. The logic behind models with fat tails (compared to those of the normal distribution) is to acknowledge that the large macroeconomic fluctuations, for example during the Covid pandemic, are hard to forecast and thus should be modelled through increased forecast uncertainty such that, importantly, large outliers do not adversely affect inference on model parameters.\footnote{This assumes that the outlier represents an `irregular' observation due to a shock rather than the co-evolution of macroeconomic variables.} Statistically, this is achieved in the posterior by down-weighting outliers through the error covariances.

The estimated model uses the same regression and state components as model (\ref{eq:non-centred1})-(\ref{eq:non-centred2}). However, we assume $\epsilon_t \sim N(0,\sigma^2\psi_t)$, where $\psi_t$ is distributed as $\mathcal{G}^{-1}(\eta/2,\eta/2)$ where $\mathcal{G}^{-1}$ denotes the inverse-Gamma distribution and $\eta$ the degree of freedom parameter of the t-distribution. Smaller degrees of freedom indicate fatter tails. To estimate the BSTS-t model, we leverage a mixture representation of the t-distribution for which derivations and sampling steps are detailed in appendix (\ref{bsts_t}). Treating $\eta$ as a random variable, figure (\ref{fig:post_eta}), based on the whole estimation sample shows that there is clear evidence for fatter tails, which are mostly due to the large outliers during the Covid pandemic. 

\begin{figure}[h]
     \centering
     \begin{subfigure}[b]{0.49\textwidth}
         \centering
         \includegraphics[width=\textwidth]{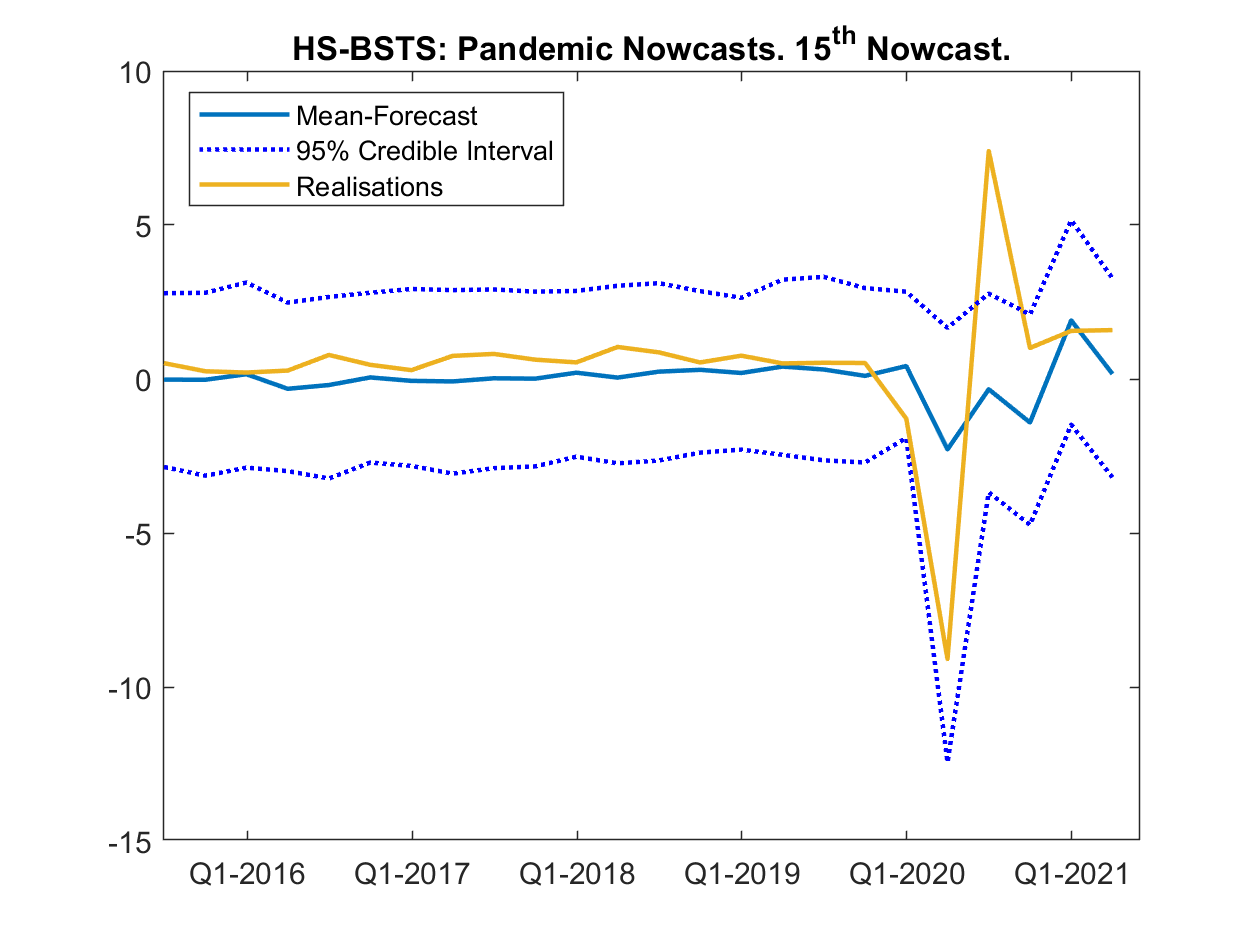}
         \label{fig:y equals x}
     \end{subfigure}
     \hfill
     \begin{subfigure}[b]{0.49\textwidth}
         \centering
         \includegraphics[width=\textwidth]{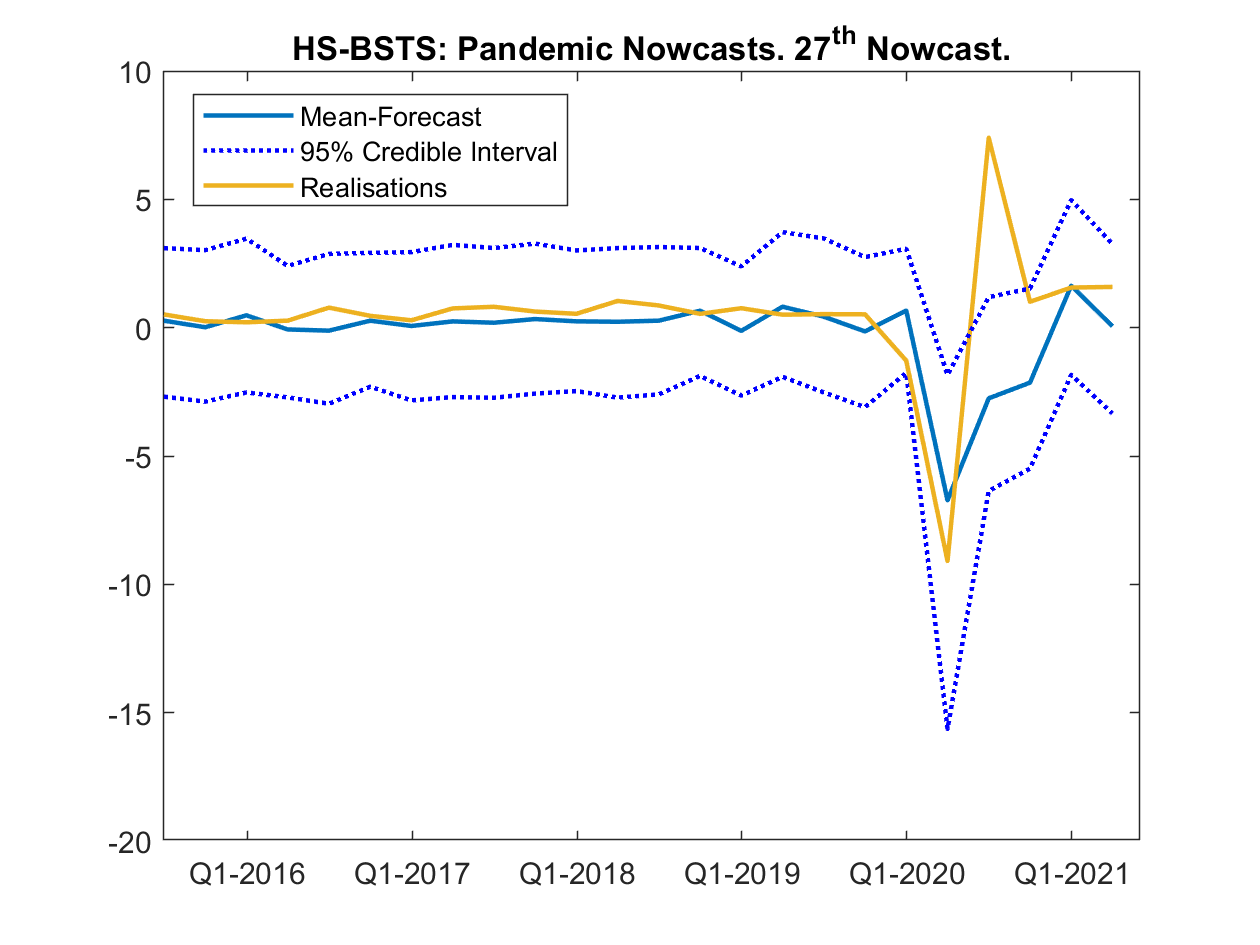}
         \label{fig:three sin x}
     \end{subfigure}
        \caption{Predictive distributions for the full HS-BSTS-t model.}
        \label{fig:pandemic_t}
\end{figure}

The nowcasts from this model (Figure \ref{fig:pandemic_t}) show that, in line with the finding of small posterior degrees of freedom for the error distribution, the forecast intervals are much wider compared to the normal BSTS models. The lower forecast interval now captures the trough during the pandemic already in the 15th nowcasting period. Surprisingly, we find that the t-model's mean prediction comes much closer to GDP growth realisation at the height of the recession in period 27. In fact, the additional nowcast period forecasts in Figure (\ref{fig:pandemic_gt}) show that already with the third publication of the Google search information within Q2 2020, we see a large downward adjustment which had not materialised in period 24 before the GT release. The posterior inclusion probabilities (\ref{fig:pip_bsts_t}) in the appendix reveal that this is because the model loads less heavily on PCE inflation of the first month (`pce2') which had not reacted much until Q2 2020, as opposed to the Google Trends and construction starts data. Yet, this new model still struggles with the upswing and presents the trade-off that even pre-Covid the predictive uncertainty is very large compared to the normal BSTS models. We believe that future research may investigate whether and how different data and modelling techniques are able to accurately forecast not only the trough, but also the peak after sharp downturns.

\section{Simulation Study} \label{sec:sim}
The empirical application above showed that the proposed BSTS models perform better in point as well as density forecasts compared to the original model of \citet{scott2014predicting} and that both the SAVS augmented horseshoe prior as well as the SSVS-BSTS exhibit a relatively sparse selection of macroeconomic data. This finding is in contrast to previous studies using macroeconomic data such as \citet{giannone2021economic} and \citet{cross2020macroeconomic} who find that priors yielding dense models generally outperform sparsity favouring priors. Since an innovation in this paper is the estimation of a latent local-linear trend which might filter out co-movement in the macroeconomic data, we compare the ability of the proposed priors to the original BSTS model \citep{scott2014predicting} in capturing both sparse and dense environments. Further, to make the simulations closer to our empirical application, we additionally test the priors’ ability to detect zero state variances. 

Specifically, we simulate local-linear-trend models as (\ref{eq:non-centred1})-(\ref{eq:non-centred2}) having either the trend variance or the local trend variance set to zero, both equal to zero, or neither equal to zero. Accordingly, we generate 20 simulated samples for $(\sigma^{\tau},\sigma^{\alpha})=\{(0.5,0),(0,0.5), (0,0),(0.5,0.5)\}$ together with either a dense or a sparse DGP, where the sparse coefficient vector is set to
\begin{equation}
    \beta_{sparse} = (1,1/2,1/3,1/4,1/5,0_{K-5})'
\end{equation}
and the dense coefficient vector is
\begin{equation}
    \beta_{dense} =
\left\{
	\begin{array}{ll}
		1/3  & \text{with probability} \; p_d \\
		0 & \text{with probability} \; 1-p_d
	\end{array}
\right.,
\end{equation}
where $p_d$ is set to 2/3. For both coefficient vectors, the dimensionality, K, is set to 300 which is high dimensional compared to the number of observations $T=150$. We account explicitly for mixed frequencies by first generating the covariate matrix according to a multivariate normal distribution with mean 0 and a covariance matrix with its $(i,j)^{th}$ element defined as $0.5^{|i-j|}$ and then skip-sample each covariate individually after the U-MIDAS methodology as in (\ref{eq:skip-sampling}). In the simulations, the true regression coefficient values as well as state variances are known; hence, we compare the performance of the different priors via coefficient bias for the regression coefficients and Dickey-Savage density ratios evaluated at zero state variances. Bias is calculated as 
\begin{equation}
    \text{Root Mean Coefficient Bias} = \sqrt{\frac{1}{20}|| \hat{\beta} - \beta||_2^2},
\end{equation}
where $\hat{\beta}$ refers to the mean of the posterior distribution. We estimate the original BSTS model with the expected model size, $\pi_0$, equal to the true number of non-zero coefficients. 

\begin{table}[]
\centering
\resizebox{0.8\textwidth}{!}{
\begin{tabular}{rcccc|cccc}
\hline
 & \multicolumn{4}{c|}{Sparse} &  \multicolumn{4}{c}{Dense} \\
 ($\sigma^{\tau},\sigma^{\alpha}$) &  (0.5,0) & (0.0.5) & (0.5,0.5) & (0,0) & (0.5,0) & (0,0.5) & (0.5,0.5) & (0,0)  \\
 \hline
 \hline
 & \multicolumn{4}{c|}{Bias} & \multicolumn{4}{c}{Bias}\\
 \hline
HS & 0.034 & 0.036 & 0.036 & 0.034 & 0.293 & 0.289 & 0.289 & 0.281\\
HS-SAVS & 0.035 & 0.035 & 0.035 & 0.035 & 0.33 & 0.327 & 0.32 & 0.321 \\
SSVS & 0.035 & 0.038 & 0.036 & 0.036 & 0.415 & 0.416 & 0.421 & 0.418\\
BSTS & 0.02 & 0.02 & 0.021 & 0.021 & 0.795 & 0.567 & 0.582 & 0.579\\
 \hline
 & \multicolumn{4}{c|}{$DS(\sigma^{\tau}=0)$} & \multicolumn{4}{c}{$DS(\sigma^{\tau}=0)$}\\
 \hline
HS  & 516.78 & 0.81 & 4.267 & 0.891  & 1.959 & 0.701 & 3.521 & 1.493 \\
SSVS  & 629.41 & 0.824 & 0.89 & 0.19 & 10.775 & 3.053 & 3.804 & 1.402 \\
 \hline
 & \multicolumn{4}{c|}{$DS(\sigma^{\alpha}=0)$} & \multicolumn{4}{c}{$DS(\sigma^{\alpha}=0)$}\\
 \hline
HS  & 0.062 & 41.587 & 722.319 & 0.026 & 0.112  & 1772.907 & 96.015 & 0.068 \\
SSVS  & 0.058 & 4.63E+10 & 1.05E+08 & 0.005 & 0.071  & 1.29E+10 & 7.56E+04 & 0.021 \\
 \hline
\end{tabular}}
\caption{Average Dickey-Savage Density ratio and bias results the simulations. Since the SAVS algorithm is performed on an iteration basis after inference, the posterior of $\sigma^{\tau,\alpha}$ remains unaffected, hence receives the same results as the HS-BSTS model.}
\label{tab:MCresults}
\end{table}

As can be seen from Table \ref{tab:MCresults}, both the non-centred BSTS models as well as the original BSTS model of \citet{scott2014predicting} do better in sparse than in dense DGPs which is similar to the finding of \citet{cross2020macroeconomic}. The largest gains of the proposed BSTS models over \citet{scott2014predicting} can be found for dense DGPs where the proposed estimators offer gains in estimation accuracy well in excess of 50\% . In sparse designs, however, the latter slightly outperforms the former. This is expected given that the spike-and-slab prior uses a point mass prior on zero and that the true expected model size is used. At the same time, it is encouraging that the differences in accuracy are very small. Among the proposed estimators in dense designs, the HS prior BSTS versions are 30-40\% more accurate compared to the SSVS-BSTS which is in line with our findings from the empirical application. Hence, these results offer the conclusion that continuous shrinkage priors are clearly preferred over spike-and-slab models in dense DGPs with a latent local-linear trend component. 

The Dickey-Savage density ratio tests confirm that the non-centred state space models are able to correctly identify which of the state variances are significant and which are not, even in high dimensional regression settings. However, the test is sensitive to correctly pinning down the regression coefficient vector: in dense designs, where the SSVS prior does worse than the horseshoe prior, the DS tests in cases (0,0.5) and (0,0) show false support for significant $\sigma^{\tau}$.\footnote{Note that we do not report DS tests for the original BSTS model. This is due to the fact that the prior on the state variance has no mass on zero and therefore is not testable.} \\

%\begin{itemize}
%    \item As the empirical study showed, the proposed extensions to the BSTS model greatly improve nowcast point as well as density fit
%    \item To test the validity of these results in larger samples, we conducted a simulation exercise which generalises the testing data sets to sparse and dense designs 
%    \item Although (reference Cross and Aubrey) have shown that GL priors are adaquate priors even for dense settings, these results have not yet been verified DGP which incorporate a latent local linear trend component. 
%    \item In particular, we generate data from $\cdots$ with the following true regression vectors [put in the betas here] 
%    \item We allow further for two different LLT variants: (1) with a significant local trend and an insignificant drift term and (2) a significant drift term, but insignificant trend term which are controlled by the %specified state standard deviations: []
%    \item List the further design specs
%    \item findings: 
%\end{itemize}

\section{Conclusion} \label{sec:conclusion}
In this paper, we investigated the added benefit of including a collection of Google Trends (GT) topics and categories in nowcasts of U.S. real GDP growth through the lens of current-generation Bayesian structural time series (BSTS) models. We extended the BSTS of \citet{scott2014predicting} to a non-centred formulation which allows shrinkage of state variances to zero in order to avoid overfitting states and therefore let the data speak about the latent structure. We further extended and compared priors used for the regression part which are agnostic about the underlying model dimensions to accommodate both sparse and dense solutions, as well as the widely successful horseshoe prior of \citet{carvalho2010horseshoe}. To make the posterior of the horseshoe prior interpretable, we applied sparsifying algorithms borrowed from the machine learning literature, which improve upon the excellent fit of the horseshoe prior itself.

We find that Google Trends improve point as well as density nowcasts in real time within the sample under investigation, where largest improvements appear prior to publication of macroeconomic information. This finding is robust across all considered models. The highest posterior inclusion probability for prediction of GDP growth across all models is obtained with the Google topics/categories `business news' and `investing'. The time-series dynamics and model impact of these GTs suggest that they provide timely signals of economic anxiety and wealth effects, respectively. Structural implications of this finding may be investigated with larger Google Trend samples and for other countries. The superior performance of the proposed models over the original BSTS model is confirmed in a simulation study which shows that among the proposed models, the horseshoe prior BSTS performs best and the largest gains in estimation accuracy can be expected in dense DGPs. It is further confirmed that the non-centred state priors are able to correctly identify the latent structure, however they are sensitive to the efficacy of the regression prior to detect signals from noise.

Finally, we applied our models to the Covid-19 pandemic period and find that Google Trends information help characterise the uncertainty during the Covid recession and subsequent recovery period. An extension of the BSTS model to student-t errors is also shown to benefit the timeliness of the forecast revisions to the changes in the macroeconomic data.  

Our work suggests some important avenues for future research. An aspect which remained unexplored in this study is that Google Trends might have time varying importance in relationship to the macroeconomic variable under investigation, as highlighted by \citet{koop2019macroeconomic}. Search terms can be highly contextual and might therefore be able to predict turning points in some periods but not in others. While, given the limited quarterly observations of Google Trends, our current investigation of this research question is somewhat limited, this will improve in significance over time. Also, nowcasting in contexts where the design is partly dense and partly sparse is a challenging problem. Our work sheds some light on this question, but it also motivates further research in this direction.

%\begin{itemize}
%    \item Summary of contributions
%    \begin{itemize}
%        \item Investigated the added benefit of using Google Trends to nowcast U.S. real GDP growth
%        \item Extensions and improvements over the conventional BSTS model:
%        \begin{itemize}
%            \item Non-centred state space to allow for state selection \& prevent overfitting
%            \item N-IG prior of Ishwaran \& Rao (2003) adapted to the BSTS
%            \item Horseshoe prior adaptation to BSTS
%            \item SAVS extension to BSTS for interpretability
%            \item provide state of the art computation algorithms
%        \end{itemize}
%    \end{itemize}
%    \item Findings
%    \begin{itemize}
%        \item GT improve point as well as density nowcasts 
%        \item Only slight improvement of GT in the first time they are released in the %HS-BSTS model
%        \item GT with highest posterior inclusion probability can be thought of as an early %warning signal: akin to financial variables
%        \item somewhat natural that GT are not a good predictor in recent times due to %relative macroeconomic tranquility
%    \end{itemize}
%    \item future avenues for research
%    \begin{itemize}
%        \item Due to time varying importance of GT, a dynamic model averaging technique of %Koop \& Onerante or TVP large regression could be an avenue
%        \item GT may only influence parts of the response's predictive distribution: QR (reference to our paper)
%        \item Larger GT database as in Ferrara et al. (2019)
%    \end{itemize}
%\end{itemize}

\pagebreak

%%TC:ignore
\bibliographystyle{chicago}
%\addcontentsline{toc}{chapter}{Bibliography}
%\pagestyle{ref}
\bibliography{reference.bib}
%\pagebreak
%%TC:endignore
\pagebreak
\appendix 
\section{Appendix} \label{sec:appendix}
\subsection{Posteriors} \label{subsec:posteriors}
In this section of the appendix, we provide the conditional posterior distributions for the regression parameters which complete the sampling steps for the Gibbs sampler detailed in (\ref{gibbssampler}).
\subsubsection{Horseshoe Prior} \label{subsubsec:hs}
Starting from model \ref{eq:non-centred1} and assuming that the states and state variances have already been drawn in steps 1.-3. in \ref{gibbssampler} which is further described in (\ref{subsec:estimation}) below. We subtract off $\tau$ such that $y-\tau = y^* = X\beta+\epsilon, \; N(0,\sigma^2I_{T})$. Printing the prior here again for convenience:
\begin{equation}
    \begin{split}
        \beta_j | \lambda_j, \nu, \sigma & \sim N(0,\lambda_j^2\nu^2\sigma^2), \;\; j \in {1,\cdots,K} \\
        \lambda_j & \sim C_{+}(0,1) \\
         \nu & \sim C_{+}(0,1) \\
        \sigma^2 & \propto \sigma^{-2}d\sigma^2.
    \end{split}
\end{equation}
Then, by standard calculations (see \citet{bhattacharya2016fast}):
\begin{equation}
    \begin{split}
        \beta | y*,\lambda, \nu, \sigma & \sim N(A^{-1}X'y^*,\sigma^2A^{-1}) \\
        A & = (X'X + \Lambda^{-1}_{*}) \\
        \Lambda_{*} & = \nu^2\text{diag}(\lambda_1^2,\dots,\lambda_K^2)
    \end{split}
\end{equation}
Instead of computing the large dimensional inverse $A^{-1}$, we rely on a data augmentation technique introduced by \citet{bhattacharya2016fast}. This reduces the computational complexity from $\mathcal{O}(K^3)$ to $\mathcal{O}(T^2K)$. Suppose the posterior is normal $N_K(\mu,\Sigma)$ with
\begin{equation}
    \Sigma = (\phi'\phi + D^{-1})^{-1}, \; \; \mu = \Sigma\phi'\alpha,
\end{equation}
where $\alpha \in \mathcal{R}^{T \times 1}$, $\phi \in \mathcal{R}^{T \times K}$ and $\text{D} \in \mathcal{R}^{K \times K}$ is symmetric positive definite. \citet{bhattacharya2016fast} show that an exact sampling algorithm is given by:
\begin{algorithm}
\caption{Fast Horseshoe Sampler}
\begin{algorithmic}[1]
    \State  Sample independently $u \sim N(0,D)$ and $\delta \sim N(0,I_T)$
     \State Set $\xi=\Phi u + \delta$
    \State Solve $(\Phi D \Phi' + I_T)w=(\alpha - \xi)$
    \State Set $\theta = u + D\Phi'w$
\end{algorithmic}
\end{algorithm} \\
Notice that $\phi = X/\sigma$, $\text{D}=\sigma^2\Lambda_*$ and $\alpha = y^*/\sigma$.

The conditional posterior for the error variance is standard \citep{makalic2015simple}:
\begin{equation}
        \sigma^2|y^{*},\beta, \lambda, \nu  \sim \mathcal{G}^{-1}((T-K)/2,(y^*-X\beta)'(y^*-X\beta)/2 + \beta'\Lambda^{-1}_*\beta/2),
\end{equation}
where $\mathcal{G}^{-1}$ denotes the inverse gamma distribution.

The posteriors of the scales $(\nu,\lambda)$ are non-standard and need different treatment. We follow \citet{polson2014bayesian} who propose an efficient slice sampler. In particular, define $\eta_j = 1/\lambda_j^2$ and $\mu_j = \beta_j/(\sigma\nu)$. Then, the conditional posterior distribution for $\eta_j$ takes the following form: 
\begin{equation}
    p(\eta_j|\nu, \sigma, \mu_j) \propto exp(-\frac{\mu_j^2}{2}\eta_j)\frac{1}{1+\eta_j}.
\end{equation}
The slice sampler is then implemented as follows:
\begin{enumerate}
    \item Sample $u_j|\eta_j$ uniformly in the interval $(0,1/(1+\eta_j))$
    \item Sample $\eta_j|\mu_j,u_j \sim exp(2/\mu_j^2)$, truncated to have zero probability outside $(0,(1-u_j)/u_j)$. 
\end{enumerate}
Now, transforming back to $\lambda_j$ yields a direct draw from its posterior distribution and by setting $\eta = 1/\nu^2$ and replacing $\mu^2_j = \sum \beta_j^2/2$ yields an equivalent draw from the conditional posterior of $\nu$. The advantages of the slice sampling algorithm include: its simplicity; that it involves no rejections; and that it requires no external parameters to be set.

\subsubsection{SSVS Prior} \label{subsubsec:ssvs}
Conditioning on the states as in \ref{subsubsec:hs}, we apply the prior:
\begin{equation}
    \begin{split}
    \beta_j|\gamma_j,\delta_j^2 & \sim \gamma_jN(0,\delta_j^2) + (1-\gamma_j)N(0,c\times\delta_j^2) \\
        \delta_j^2 & \sim \mathcal{G}^{-1}(a_1.a_2) \\
        \gamma_j & \sim Bernoulli(\pi_0) \\
        \pi_0 & \sim \mathcal{B}(b_1,b_2),
    \end{split}
\end{equation} 
where $\mathcal{B}$ stands for the beta distribution. The conditional posteriors are standard and derived for example in \citet{george1993variable} and \citet{ishwaran2005spike}. The difference to the prior of \citet{george1993variable} lies in the additional prior for $\delta^2_j$ which is assumed to be inverse gamma. It can be shown that this implies a mixture of student-t distributions for $\beta_j$ marginally \citep{konrath2008bayesian}. We sample from the conditional posteriors in the following way:
\begin{algorithm}
\caption{SSVS Sampler}
\begin{algorithmic}[1]
    \State  For $j \in \{1,\cdots,K\}$, sample each $\gamma_j|\beta_j,\delta^2_j,\pi_0,y \sim (1-\pi)N(\beta_j|0,c \times\delta_j^2)I_{\gamma_j=0}$ $+ \pi_0N(\beta_j|\beta_j|0,\delta_j^2)I_{\gamma=1}$
     \State Sample $\pi_0 \sim \mathcal{B}(b_1 + n_1, b_2 + K-n_1)$, where $n_1 = \sum_jI_{\gamma_j=1}$
    \State Sample $\beta|\gamma,\delta^2,\sigma^2,y \sim N(A^{-1}X'y^*/\sigma^2,A)$, where $A^{-1} = X'X/\sigma^2 + D{-1}$, $D = \text{diag}(\delta_j^2\gamma_j)$
    \State Sample $\sigma^2 \sim \mathcal{G}^{-1}(\overline{c},\overline{C})$, where $\overline{c} = \underline{c} + \frac{T}{2}$, $\overline{C} = \underline{C} + \frac{1}{2}((y^*-X\beta)'(y^*-X\beta))$ and $p(\sigma^2) \sim \mathcal{G}(\underline{c},\underline{C})$
\end{algorithmic}
\end{algorithm} \\

\subsection{Robustness Check for the Horseshoe Prior}\label{robustness}
The Cauchy distribution can be a challenging distribution to fit. Due to no analytically available moments, a posterior distribution in which the prior information dominates the likelihood, Cauchy priors might induce vanishing posterior moments \citep{ghosh2018use,piironen2017sparsity}. \citet{piironen2017sparsity} provide a way in which potential problems due to weak identification can be diagnosed, which is based on the prior induced effective model size distribution. 

Assuming a scale mixture of normal prior such as (\ref{eq:glpriors}), the conditional posterior $p(\beta|\Lambda_*,\tau,\sigma^2,y)$ within a normal linear regression can be written as: 
\begin{equation}
    \overline{\beta}_j = (1-\kappa_j)\hat{\beta}_j
\end{equation}
where
\begin{equation} \label{eq:kappa}
    \kappa_j = \frac{1}{1+T\sigma^{-2}\nu^2s_j^2\lambda^2_j},
\end{equation}
and $\overline{\beta}_j$ and $\hat{\beta}_j$ refer to the mean of the posterior for $\beta_j|\bullet$ and the maximum likelihood solution respectively. $\kappa_j$ can be regarded as a shrinkage coefficient as it is defined over the unit interval and therefore dictates how much shrinkage is applied to the maximum likelihood solution. It is easy to verify that when $\lambda_j\nu \rightarrow \infty$, then $\kappa_j \rightarrow 0$ and when $\lambda_j\nu \rightarrow 0$, then $\kappa_j \rightarrow 1$. The distribution of the shrinkage coefficient $p(\kappa_j|\Lambda_*,\nu,\sigma^2,y)$ is implicitly defined through the priors for the hyperparameters ($\Lambda,\nu$). By applying the change of variable theorem, it can be shown that for the horseshoe prior, this distribution is proportional to $\mathcal{B}(0.5,0.5)$. In fact, by plotting this distribution alongside the distribution implied by a spike-and-slab prior with zero point-mass spike and slab with infinite scale, the horseshoe prior provides a continuous approximation to the SSVS.

\begin{figure}[h]
    \centering
    \includegraphics[width=\textwidth]{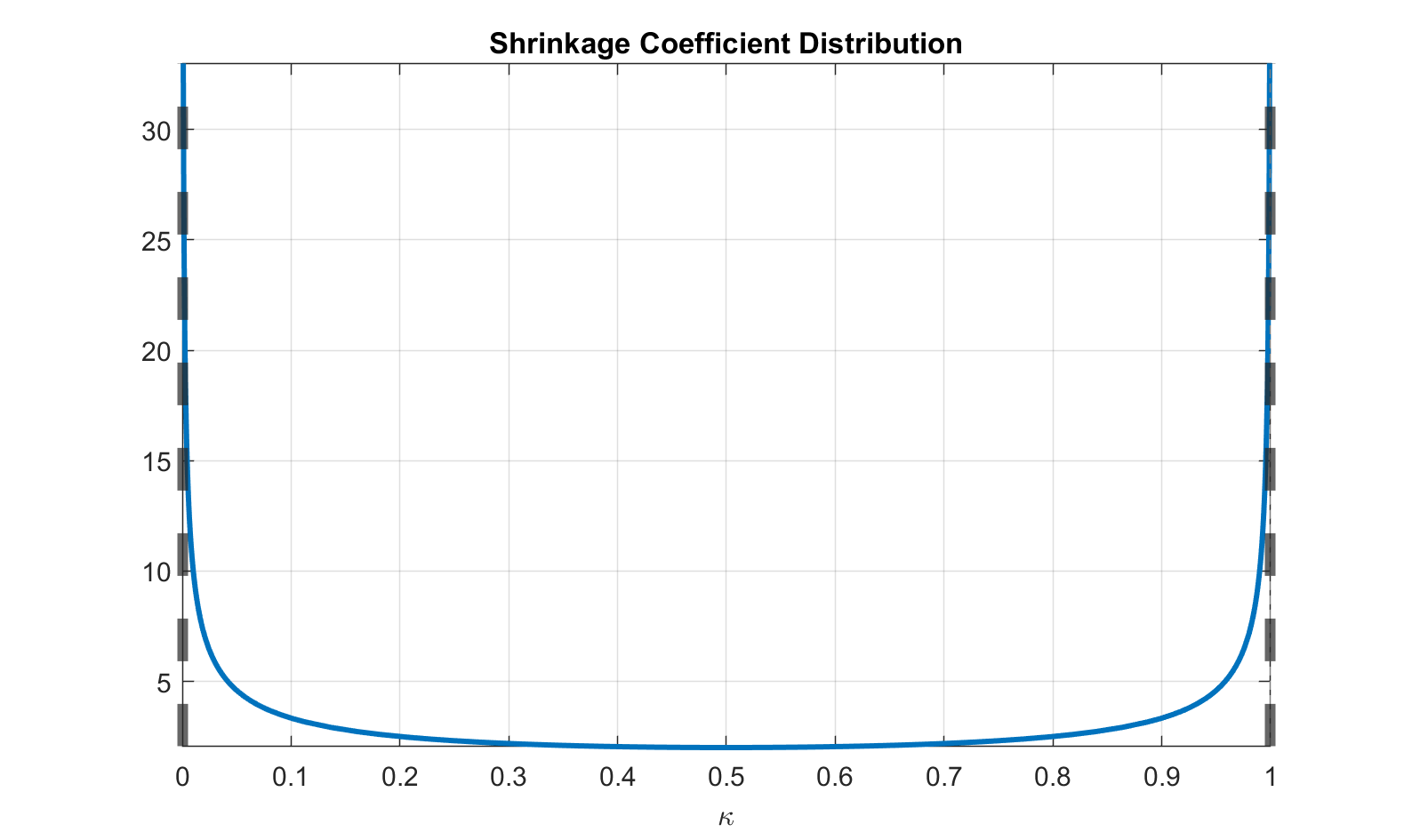}
      \caption{$\kappa|\bullet$ distributions for the horseshoe prior (blue lines) and the spike-and-slab prior (dashed grey lines) with zero point mass and infinite slab scale.}
    \label{fig:prior_predictive}
\end{figure}

Now, by summing over the shrinkage coefficients in (\ref{eq:kappa}), the authors provide a measure of ``active regressors'': $ m_{eff} = \sum (1-\kappa_j)$, i.e. large slopes, which are used to illicit a prior for the global shrinkage scale. Due to the aggressive shrinkage profile of the horseshoe prior, the distribution over $m_{eff}$ can be thought of as an effective model size distribution.

\citet{piironen2017sparsity} show that, drawing from the prior predictive distribution with the standard horseshoe hierarchy, the prior effective model size distribution diverges to the largest possible number of slopes. This is confirmed in Figure (\ref{fig:prior_predictive}). 

\begin{figure}[h]
    \centering
    \includegraphics[width=0.7\textwidth]{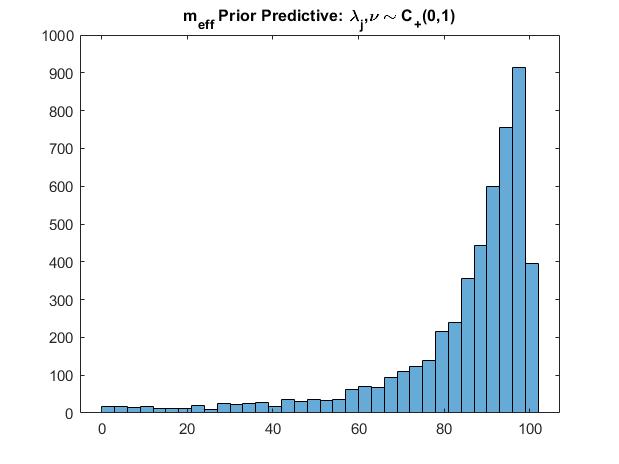}
      \caption{$m_{eff}$ generated by setting $T=K=100$ and $\sigma=1$. Distribution based on 5000 samples according to the prior in (\ref{eq:glpriors}). }
    \label{fig:prior_predictive}
\end{figure}

Since, as discussed above, the prior predictive distribution of $m_{eff}$ diverges to K, weak identification of either the local or global scale parameters would lead to a divergent posterior $m_{eff}$ distribution, which we use as a diagnostic tool.

However, some caveats must be highlighted: (1) the number of active coefficients formulation is derived from the \textit{conditional} posterior, and hence does not account for uncertainty in ($\nu,\sigma^2$); and (2) it makes further two strong assumptions: the covariates are uncorrelated and the maximum likelihood solution to $\beta$ exists. Hence, caution should be exercised in taking this approach at face value with high-dimensional macro data, which have typically large cross-sectional correlation.

\begin{figure}[h]
     \centering
     \begin{subfigure}[b]{0.49\textwidth}
         \centering
         \includegraphics[width=\textwidth]{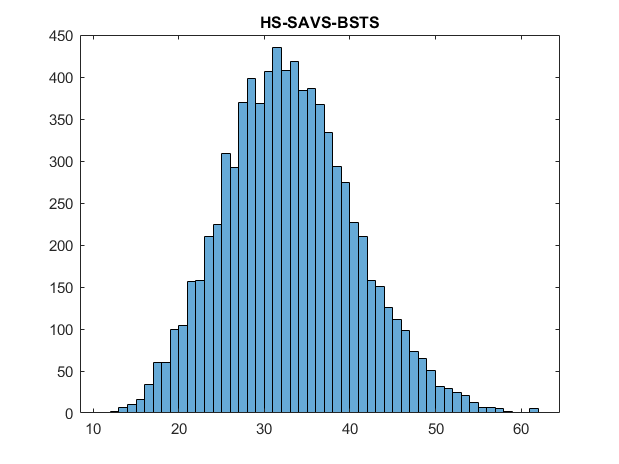}
     \end{subfigure}
     \hfill
     \begin{subfigure}[b]{0.49\textwidth}
         \centering
         \includegraphics[width=\textwidth]{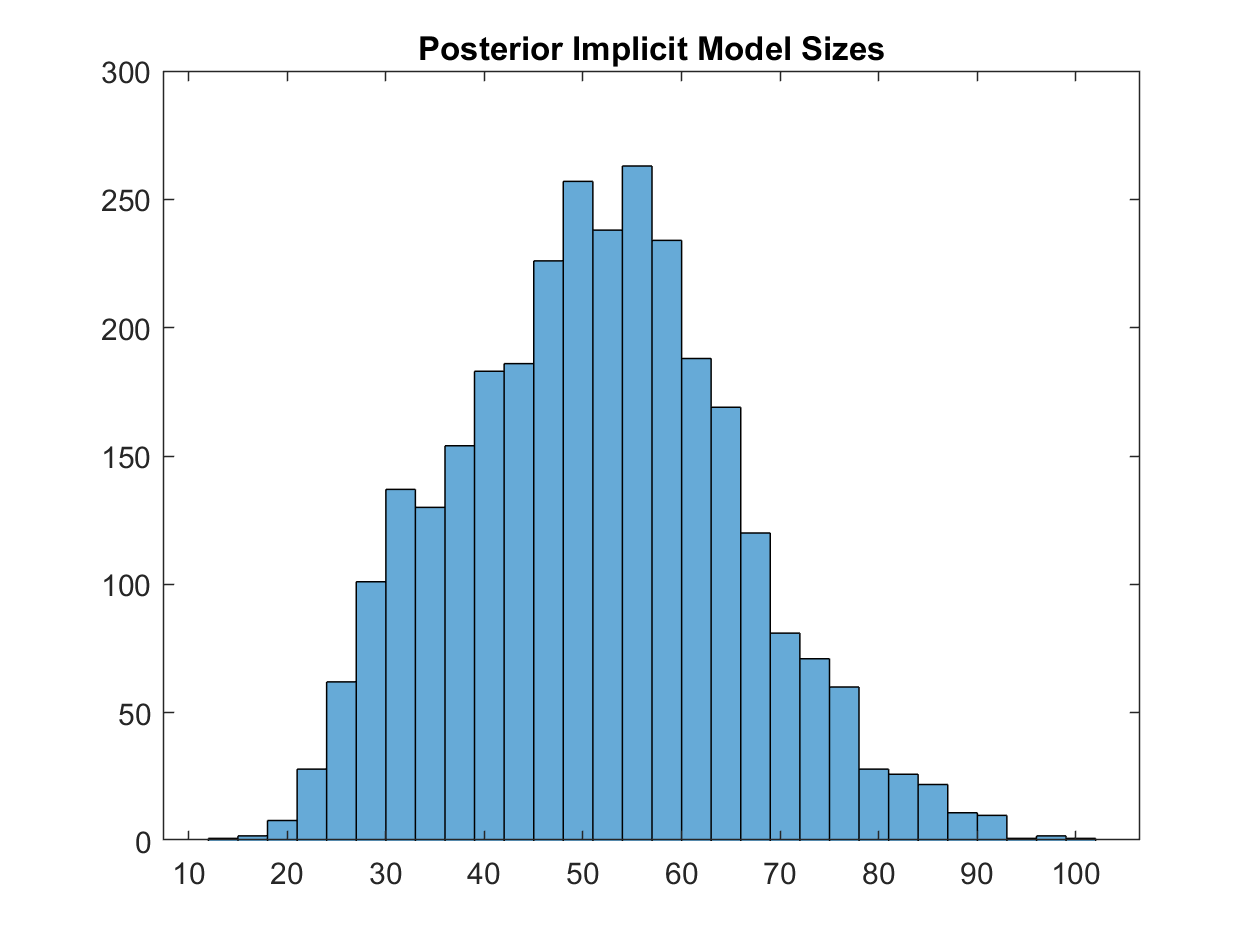}
     \end{subfigure}
    
        \caption{Posterior model size distribution for the HS-SAVS-BSTS (left) and the posterior of $m_{eff}$ for the HS-BSTS (right)}
        \label{fig:modelsizes}
\end{figure}

Nevertheless, we provide in Figure \ref{fig:modelsizes} the posterior effective model size distributions for the HS-BSTS model based on the entire sample without ragged edges. As the figure clearly shows, the effective model size is not divergent to K and is very similarly distributed to the SAVS model size distribution. Hence, for the HS-BSTS model, there is no indication of weakly identified posterior scale processes.

\subsection{ARMA vs LLT Estimation}
While the nowcasting exercise has demonstrated the statistical support for a local-linear-trend model, and hence support for shifts in the long-run rate of GDP growth, it is an important question as to whether GDP dynamics within a high-dimensional regressor setting is better modelled via ARMA components. 

Hence, we also estimate the regression as in the paper, but with ARMA components instead of a LLT. The model considered is:

\begin{equation}
    \begin{split}
        y_t & = \beta_0 + x_t'\beta + \sum^{p}_{j=1}\rho_jy_{t-j} + \epsilon_t \\ 
        \epsilon_t & = u_t + \sum^q_{m=1}\psi_mu_{t-q},
    \end{split}
\end{equation}
where $|\psi|<1$ for identification purposes and $\{u_t\}_0^T \sim N(0,\sigma^2)$.

The posterior for the AR coefficients is modelled via the horseshoe prior. We further assume a uniform prior on the interval (-1,1) for $\psi \sim \mathcal{U}(-1,1)$. To estimate the ARMA components, we follow \citet{chan2017notes} by using a band and sparse matrix representation which allows for very fast computation by avoiding recursive algorithms. The order of the ARMA components are chosen via the best out-of-sample performance.

\begin{figure}[H]
    \centering
    \includegraphics[width=\textwidth]{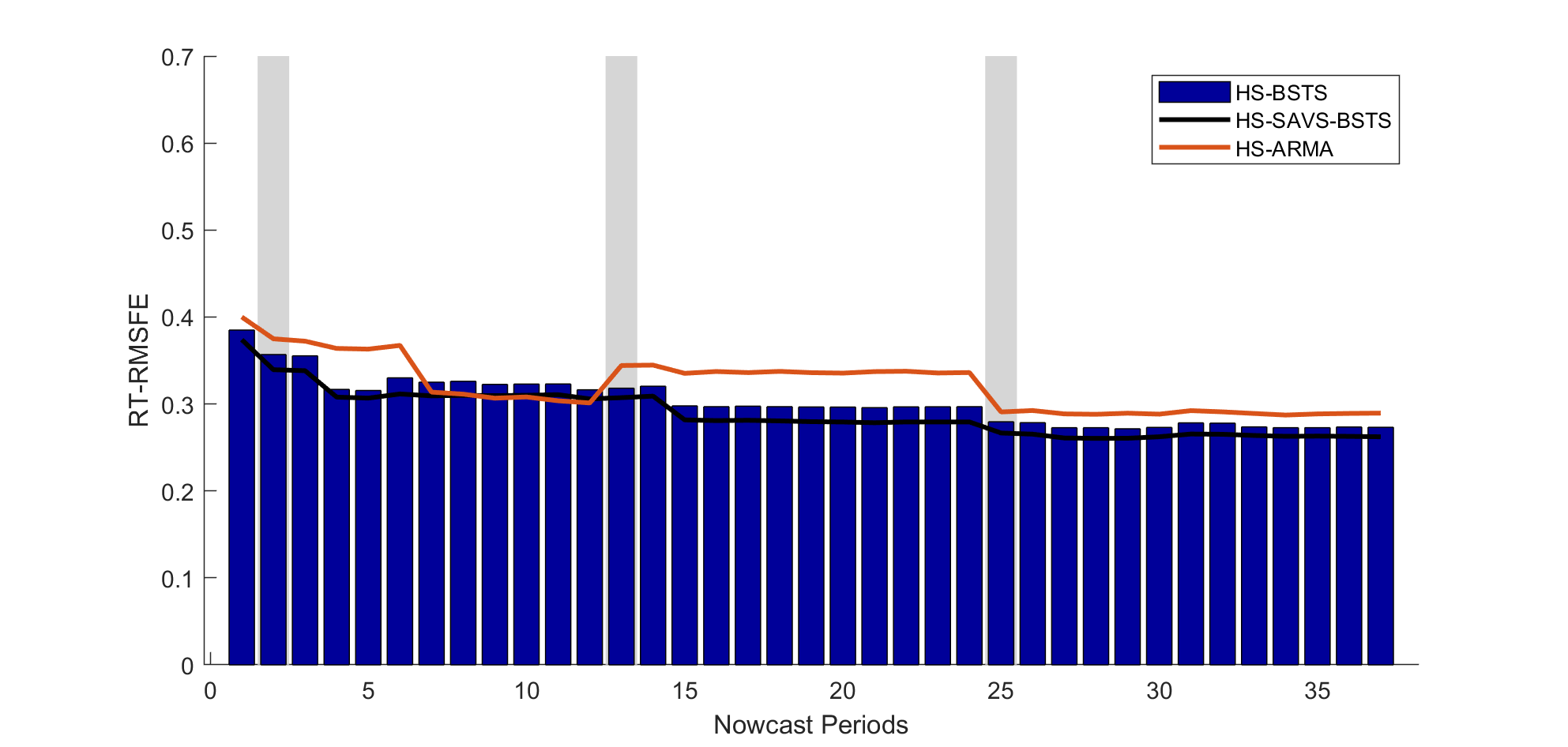}
      \caption{Real-Time-RMSFE for 37 nowcast periods.}
    \label{fig:rmsfe_arma}
\end{figure}

\begin{figure}[H]
    \centering
    \includegraphics[width=\textwidth]{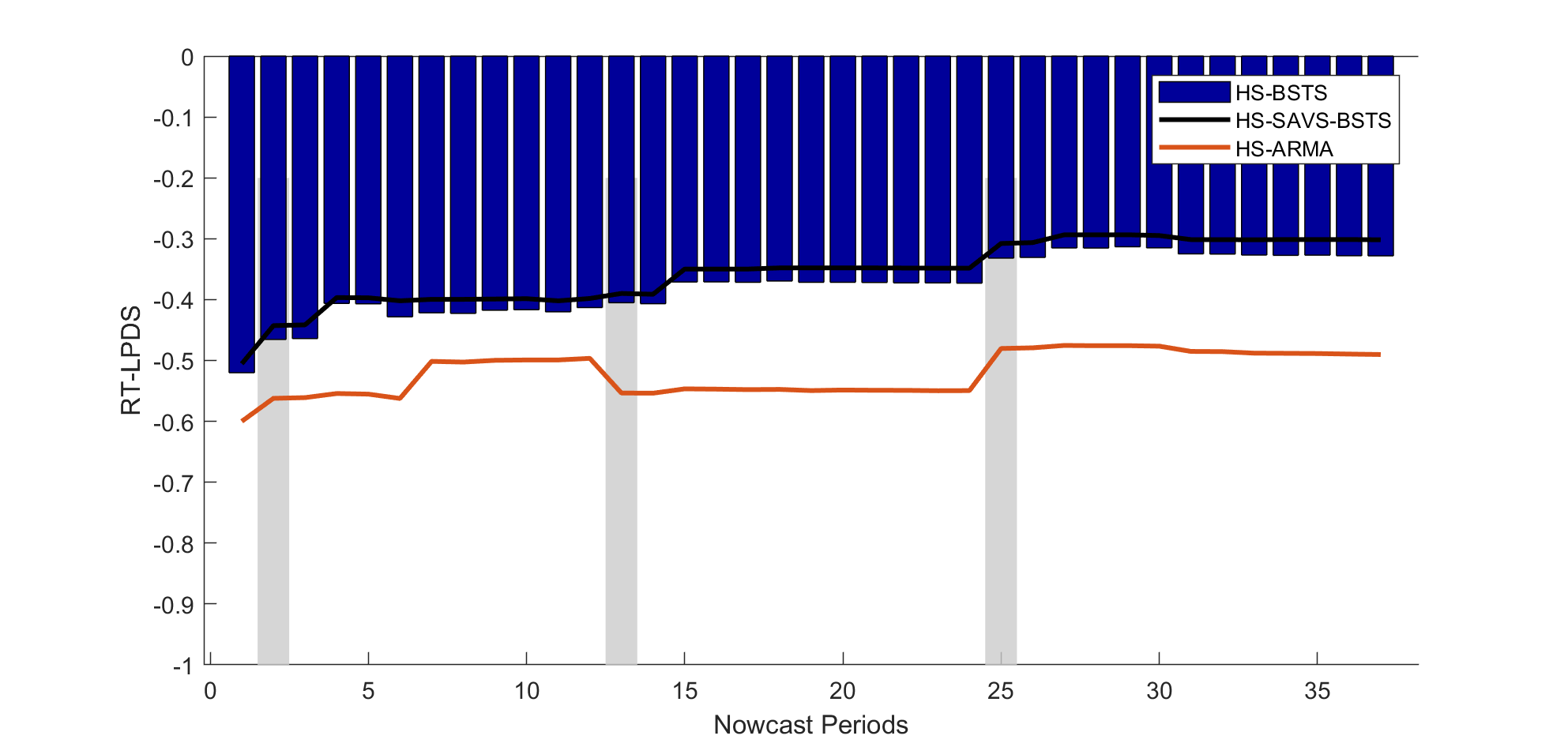}
      \caption{Real-Time-LPDS for 37 nowcast periods.}
    \label{fig:lpds_arma}
\end{figure}

\begin{figure}[H]
    \centering
    \includegraphics[width=\textwidth]{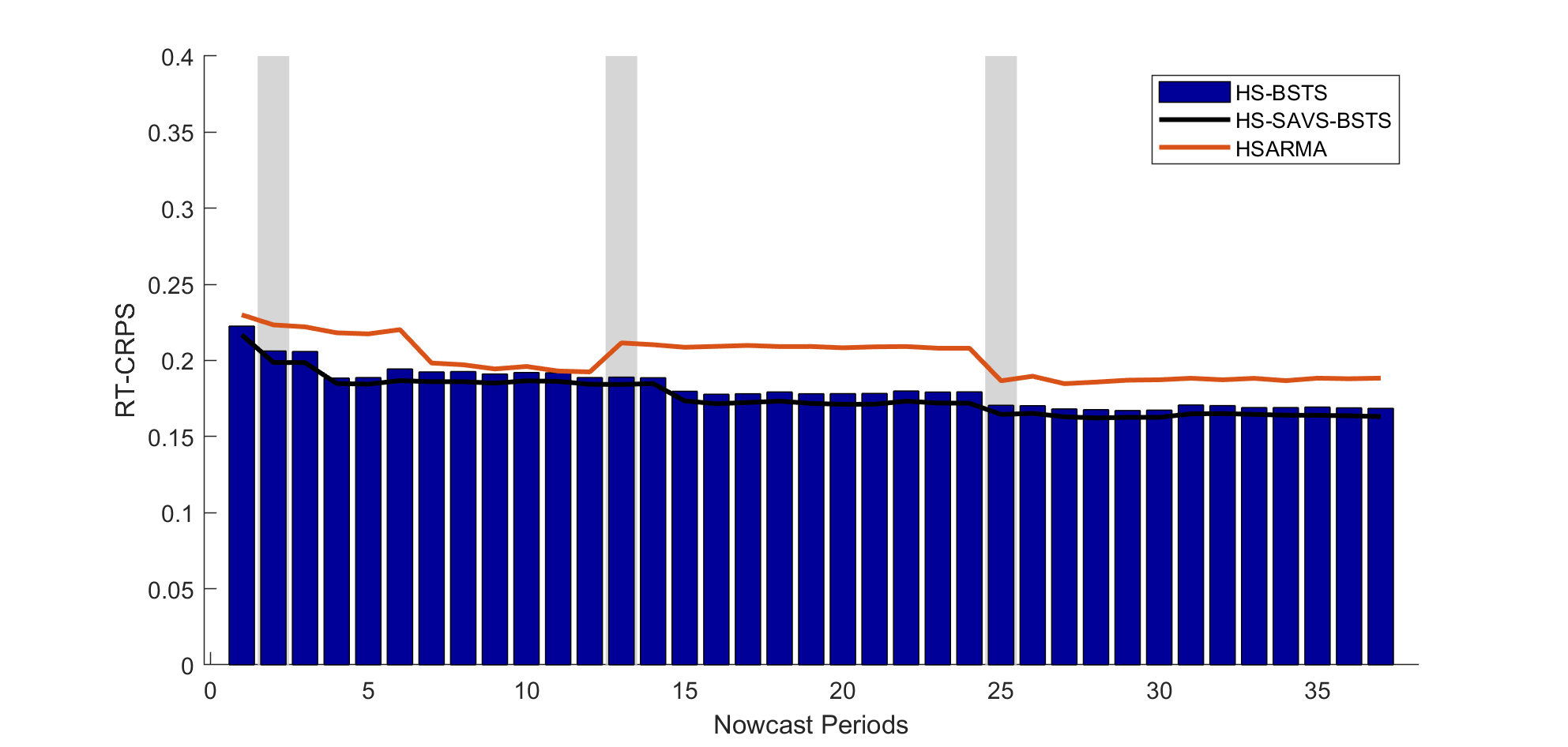}
      \caption{Real-Time-CRPS for 37 nowcast periods.}
    \label{fig:crps_arma}
\end{figure}

As can be clearly seen from Figures \ref{fig:rmsfe_arma}-\ref{fig:crps_arma}, the point as well as density nowcasts are clearly worse for the ARMA compared to the LLT horseshoe prior models. 

\begin{figure}[h]
    \centering
    \includegraphics[width=\textwidth]{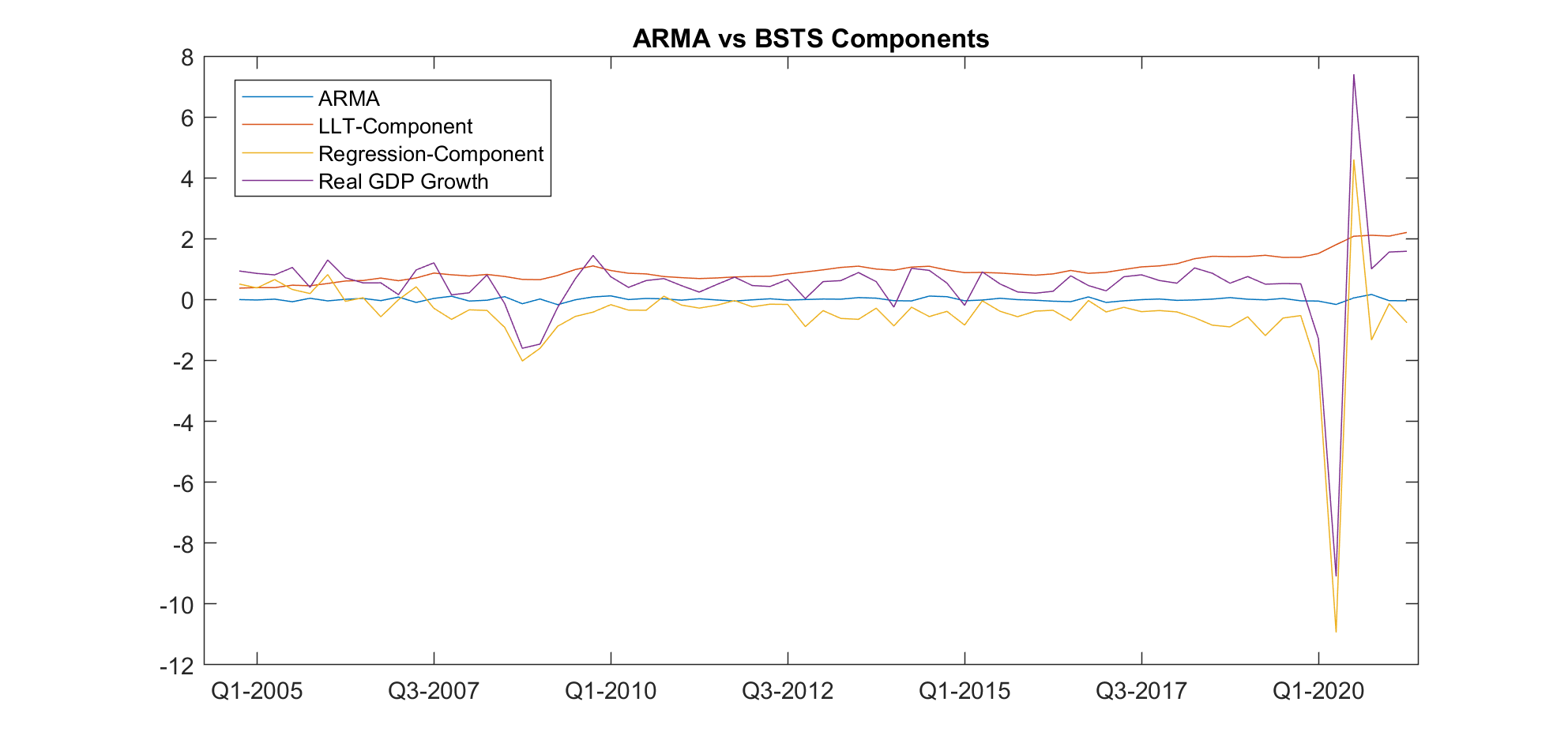}
      \caption{Trend and regression decomposition for the HS-BSTS model and the best performing HS-ARMA model (ARMA(1,1)).}
    \label{fig:bsts_arma}
\end{figure}

This superior performance is related to the argument made above: the LLT model captures the slow-moving long-run growth component which gives the BSTS models the flexibility to capture deviations from this trend gathered from the explanatory information. This is underlined in figure (\ref{fig:bsts_arma}): the LLT component captures smooth transitions in GDP growth, while the large data set captures large troughs and peaks, such as during the financial crisis. By contrast, the ARMA component is of very small magnitude and displays erratic short-run movements. This suggests that short run dynamics are indeed better captured by the macro and Google Trend information.

\subsection{State Space Estimation and Forecasting}
\subsubsection{Estimation} \label{subsec:estimation}
Assume analogously to \ref{subsubsec:hs} and \ref{subsubsec:ssvs} that all regression parameters have been sampled such that conditionally on $\beta$, we estimate $y-X\beta = \hat{y}_t =  \tau_0 + \sigma_{\tau}\tilde{\tau_t} + t\alpha_0 + \sigma_{\alpha}\sum^t_{s=1}\tilde{\alpha_t} + \epsilon, \; \epsilon_t \sim N(0,\sigma_y) $ and $\tilde{\tau}_t = \tilde{\tau}_{t-1} + u^{\tau}_t, \; u^{\tau}_t \sim N(0,1)$, $\tilde{\alpha}_t = \tilde{\alpha}_{t-1} + u^{\alpha}_t, \; u^{\alpha}_t \sim N(0,1)$. Since the state processes $\{\tilde{\tau},\tilde{\alpha}\}_{t=1}^{T}$ are independent of the other parameters in the non-centred fomrulation, we proceed by first estimating the states and then $\theta  = \{\tilde{\tau_0},\tilde{\alpha},\sigma_{\tau},\sigma_{\alpha}\}$. \\
States $\{\tilde{\tau},\tilde{\alpha}\}_{t=1}^{T}$ can be sampled by any state space algorithm, e.g. \citet{durbin2002simple}, \citet{carter1994gibbs} or \citet{fruhwirth1994data}. We instead opt for the precision sampler by \citet{chan2017notes} which exploits the joint distribution of the states which paired with sparse matrix operations yields significant increases in statistical as well as computational efficiency \citep{grant2017bayesian}. Since $\tilde{\alpha_s}$ enters in the observation equation as a sum, we define $\tilde{A}_t = \sum_{s=1}^t\tilde{\alpha}_s$. Notice that equation (\ref{eq:non-centred2}) implies that $\boldsymbol{H}\boldsymbol{\tilde{\alpha}}=\boldsymbol{\tilde{u}^{\alpha}}$, where $\boldsymbol{H}$ is the first difference matrix\footnote{[Put in here the first difference matrix as defines on p.81, Chan (2017)]} and $\boldsymbol{\tilde{u}^{\alpha}}\sim N(\boldsymbol{0},\boldsymbol{I_T})$. Notice that $\tilde{A}_t = \tilde{\alpha}_1$ which implies that $\tilde{A}_{t}-\tilde{A}_{t-1} = \tilde{\alpha}_t$. Hence, this gives us back the desired $\boldsymbol{H}\boldsymbol{\tilde{A}}=\boldsymbol{\tilde{\alpha}}$. Solving $\boldsymbol{\tilde{A}} = \boldsymbol{H}^{-1}\boldsymbol{\tilde{u}^{\alpha}}= \boldsymbol{H^{-2}}\boldsymbol{\tilde{u}^{\alpha}}$. Therefore
\begin{equation}
    \boldsymbol{\tilde{A}} \sim N(\boldsymbol{0},(\boldsymbol{H^2}\:'\boldsymbol{H^2})^{-1})
\end{equation}
To sample the states jointly, define $\xi = (\tilde{\boldsymbol{\tau}}',\tilde{\boldsymbol{A}}')'$. Then $\hat{y}$ can be re-written as:
\begin{equation}\label{eq:model}
    \hat{y} = \tau_0\boldsymbol{1}_T + \alpha_0\boldsymbol{1}_{1:T} + X_{\xi}\xi + \epsilon,
\end{equation}
where $\boldsymbol{1}_{1:T}$ is defined as $(1,2, \cdots, T)'$ and $\boldsymbol{X}_{\xi} = (\sigma_{\tau}I_T, \sigma_{\alpha}I_T)$. Since $\boldsymbol{X}_{\xi}$ is a sparse matrix, manipulations in programs which utilise sparse matrix operations will be very fast.\\
Similar calculations result in the implicit prior $\tilde{\boldsymbol{\tau}} \sim N(\boldsymbol{0},(\boldsymbol{H}'\boldsymbol{H})^{-1})$. Now, since by assumption $\boldsymbol{\tau}$ and $\boldsymbol{\tilde{A}}$ are independent, the combined for $\xi$ is:
\begin{equation}
    \xi \sim N(0,\boldsymbol{P}_{\xi}^{-1}),
\end{equation}
where $\boldsymbol{P}_{\xi} = \text{diag}(\boldsymbol{H}'\boldsymbol{H},\boldsymbol{H}^2{}'\boldsymbol{H}^2)$. The posterior is thus standard:
\begin{equation}
    p(\xi|\boldsymbol{y},\sigma^2_y) \sim N(\overline{\xi},A^{-1}_{\xi})
\end{equation}
where $K_{\xi} = P_{\xi} + \frac{1}{\sigma^2_y}\boldsymbol{X}'_{\xi}\boldsymbol{X}_{\xi}$ and $\overline{\xi} = \boldsymbol{K}^{-1}_{\xi}(\frac{1}{\sigma^2_{y}}\boldsymbol{X}_{\xi}'(\boldsymbol{y}-\tau_0\boldsymbol{1}_{T}-\alpha_0\boldsymbol{1}_{1:T}))$. \\
Conditionally on $\xi$, $\theta$ are drawn by simple linear regression results, where we specify a generic prior covariance as $V_{\theta} = \text{diag}(1,1,0.1,0.1)$ and prior mean $\theta_0 = (0,0,0,0)$. In particular, define $X_{\theta} = (\boldsymbol{1}_T,\boldsymbol{1}_{1:T},\tilde{\tau},\tilde{\boldsymbol{A}})$. Then: 

\begin{equation}
    \begin{split}
        \theta|\boldsymbol{y},\sigma_y^2 & \sim N(\overline{\theta},A^{-1}_{\theta}) \\
        A_{\theta} & = (V^{-1}_{\theta} + \frac{1}{\sigma^2_y}X'_{\theta}X_{\theta}) \\
        \overline{\theta} & = A^{-1}_{\theta}(V^{-1}_{\theta}\theta_0+\frac{1}{\sigma^2_y}X'_{\theta}\hat{y}).
    \end{split}
\end{equation}

\subsubsection{Forecasting} \label{forecasting}
Taking equation (\ref{equ:bsts}) as our starting point, it is well known that the predictive density $p(y_t|\boldsymbol{y^{t-1}},\beta,\theta,\sigma^2_y)$, where $\boldsymbol{y^{t-1}}=(y_1,\cdots,y_{t-1})$, can be generated by the Kalman filter. Since the state space is instead estimated by precision sampling, and hence, without Kalman recursions, the literature has proposed (1) conditionally optimal Kalman mixture approximations \citep{bitto2019achieving}, (2) pure simulation based methods to approximate (1) \citep{belmonte2014hierarchical}, and (3) what \cite{bitto2019achieving} call naive Gaussian mixture approximation (see A.1.2.2 of \cite{bitto2019achieving}). In simulations as well as the empirical example we found that results are very similar independent of the sampling technique. For computationally simplicity we present here method (2). \\
The predictive on-step-ahead distribution $p(y_t|\boldsymbol{y_{t-1}})$ can be generated by first drawing from the non-centred states which with the draws of the other model parameters yield draws from the predictive. More specifically, for posterior draw $m=1,\cdots,M$:
\begin{enumerate}
    \item Draw $(\tilde{\tau}^{(m)}_t, \tilde{\alpha}^{(m)}_t)$ from $N(\tilde{\tau}^{(m)}_{t-1},1)$ and $N(\tilde{\alpha}^{(m)}_{t-1},1)$ respectively
    \item Generate $\alpha_t = \alpha_0^{(m)} + \sigma_{\alpha}^{(m)}\tilde{\alpha}^{(m)}_t$ and $\tau^{(m)} = \tau_0^{(m)} + \sigma_{\tau}^{(m)}\tilde{\tau}_t^{(m)} + t\alpha_0^{(m)} + \sigma_{\alpha}^{(m)} \sum_{s=1}^t\alpha_s^{m}$
    \item Generate $x'_t\beta^{(m)} + \tau^{(m)} + \sigma_y^{(m)}u$, where $u \sim N(0,1)$
\end{enumerate}
To obtain an approximation to the continuous approximation to $p(y_t|\boldsymbol{y^{t-1}})$, one can then use a kernel density smoother such as "kdensity" in Matlab.

\subsection{BSTS-t Model}\label{bsts_t}
The model to be estimated is:
\begin{equation} 
    y_t = \tau_0 + \sigma_{\tau}\tilde{\tau}_t + t\alpha_0 + \sigma_{\alpha}\sum_{s=1}^t\tilde{\alpha}_t + x_t'\beta + \epsilon_t, \; \epsilon_t \sim N(0,\sigma^2\psi_t), \; \psi_t \sim \mathcal{G}^{-1}(\eta/2,\eta/2)
\end{equation}
and
\begin{equation} 
    \begin{split}
        \tilde{\tau}_t & = \tilde{\tau}_{t-1} + \tilde{u}^{\tau}_t, \tilde{u}^{\tau}_t \sim N(0,1) \\
        \tilde{\alpha}_t & = \tilde{\alpha}_{t-1} + \tilde{u}^{\alpha}_t, \tilde{u}^{\alpha}_t \sim N(0,1)
    \end{split}
\end{equation} 
Compared to the normal BSTS models,, $\boldsymbol{\psi} = (\psi_1,\cdots,\psi_T)'$ ans $\eta$ are additional unknown parameters. 

For exposition, we treat $\eta$ for now as known. The Gibbs sampler to draw inference on this model needs the conditional posteriors: $p(\beta|y\theta,\tilde{\tau},\tilde{\alpha},\sigma^2_y,\boldsymbol{\psi})$, $p(\tilde{\tau},\tilde{\alpha}|y,\beta,\theta,\sigma^2_y,\boldsymbol{\psi})$, $p(\theta|y,\tilde{\tau},\tilde{\alpha},\sigma^2_y,\boldsymbol{\psi})$ and $p(\sigma^2_y|y,\theta,\tilde{\tau},\tilde{\alpha},\boldsymbol{\psi})$.

To derive $p(\beta|\bullet)$, rewrite again $y^*=y-\tau = X\beta+\epsilon$. Then $\epsilon_t \sim (0_T,\sigma^2_y\Theta)$, where $\Theta = diag(\psi_1,\cdots,\psi_T)$. Defining $\Sigma = \sigma^2\Theta$:

\begin{equation}
\begin{split}
    \beta|\bullet & \sim N(\hat{\beta}_1,D_1) \\
    D_1 & = (\Lambda^{-1}_*+\frac{1}{\sigma^2_y}X'\Sigma^{-1}X)^{-1} \\
    \hat{\beta}_1 & = A^{-1}\frac{1}{\sigma^2_y}\Sigma^{-1}y^*,
\end{split}
\end{equation}
where the diagonal of $\Lambda_*$ is populated by the shrinkage scales of either the horseshoe prior or the SSVS.  \footnote{Since the horseshoe uses a conjugate formulation, the posterior moments for the horseshoe are $D_1=\sigma^2(X'\Sigma^{-1}X + \Lambda^{-1})$ and $\hat{\beta}_1 = D_1^{-1}X'\Sigma^{-1}y^*$.}

To sample from $p(\theta|\bullet)$, rewrite again $\hat{y} = y_t-x_t'\beta = \tau_0 + \sigma_{\tau}\tilde{\tau}_t + t\alpha_0 + \sigma_{\alpha}\sum_{s=1}^t\tilde{\alpha}_t + \epsilon_t$, $\epsilon_t \sim N(0,\sigma^2_y\psi_t)$ and $\psi_t \sim \mathcal{G}^{-1}(\eta/2,\eta/2)$. Again, define  $X_{\theta} = (\boldsymbol{1}_T,\boldsymbol{1}_{1:T},\tilde{\tau},\tilde{\boldsymbol{A}})$. Then the posterior is:

\begin{equation}
    \begin{split}
        \theta|\bullet & \sim N(\hat{\beta}_2,D_2) \\
        D_2 & = (V^{-1}_{\theta}+\frac{1}{\sigma^2_y}X'_{\theta}\Sigma^{-1}X_{\theta})^{-1} \\
        \hat{\beta}_2 & = D_2(V^{-1}_{\theta}\theta_0 + \frac{1}{\sigma^2_y}X'_{\theta}\Sigma^2\hat{y})
    \end{split}
\end{equation}

To sample from $(\tilde{\tau},\tilde{\alpha}|\bullet)$, take equation \ref{eq:model} with $\epsilon_t \sim N(0,\sigma^2_y\psi_t)$ and $\psi_t \sim \mathcal{G}^{-1}(\eta/2,\eta/2)$. Then the posterior with the same steps as above is: 

\begin{equation}
    \begin{split}
        \xi | \hat{y},\sigma^2_y & \sim N(\hat{\beta}_3,D_3) \\
        D_3 & = (P_{\xi}+\frac{1}{\sigma^2_y}X'_{\xi}\Sigma^{-1}X_{\xi})^{-1} \\
        \hat{\beta}_3 & = \boldsymbol{K}^{-1}_{\xi}(\frac{1}{\sigma^2_{y}}\boldsymbol{X}_{\xi}'\Sigma^{-1}(\boldsymbol{y}-\tau_0\boldsymbol{1}_{T}-\alpha_0\boldsymbol{1}_{1:T})).
    \end{split}
\end{equation}

To derive the posterior $p(\boldsymbol{\psi}|\bullet)$, notice that each $\psi_t$ is univariate and independently distributed. Hence:

\begin{equation}
    p(\boldsymbol{\psi}|\bullet) \propto \prod_{t=1}^T \psi_t^{-\frac{\eta+1}{2}+1}e^{-\frac{1}{2\psi_t}\frac{(\eta+y_t-\tau-x_t'\beta)^2}{\sigma^2_y}}.
\end{equation}
Notice that these are kernels of the inverse-Gamma distribution: 

\begin{equation}
    \psi_t \sim \mathcal{G}^{-1}(\frac{\eta+1}{2},\frac{1}{2}\frac{(y_t-\tau-x_t'\beta)^2}{\sigma^2_y})
\end{equation}

Finally, regarding the unknown $\eta$, we specify a uniform prior $\eta \sim U[2,50]$. The lower limit ensures that the variance $\sigma^2_y$ exists, and 50 is chosen to be reasonably large such that the upper limit generates an error variance close to a normal. The conditional posterior boils down to:
\begin{equation}
    \begin{split}
        p(\eta|\bullet) & \propto p(\boldsymbol{\psi}p(\eta)) \\
        & \propto \prod_{t=1}^T\frac{(\eta/2)^{\frac{\eta}{2}}}{\Gamma(\eta/2)} \psi_t^{-(\frac{\eta}{2}+1)}e^{-\frac{\eta}{2\psi_t}} \\
        & = \frac{(\eta/2)^{\frac{T}{2}}}{\Gamma(\eta/2)^T}(\prod_{t=1}^T)^{-(\frac{\eta}{2}+1)}e^{\frac{\eta}{2}\sum_{t=1}^T\psi_t^{-1}}
    \end{split}
\end{equation}
where the first definition follow from the fact that the priors are independent. This distribution is non-standard. To sample from this distribution, we make use of an independent Metrolpolis-Hastings within Gibbs sampling step.

By slight abuse of notation, define the target density as f, the current state of the Markov chain as X and the proposal state as Y, then the proposal Y is accepted with probability 

\begin{equation}
    \alpha(X,Y) = min\Bigg\{\frac{f(Y)g(X)}{f(X)g(Y)},1\Bigg\},
\end{equation}
where $g(.)$ is the proposal density. In order for the Metropolis-Hastings sampler to quickly explore the typical set of $\eta|\bullet$, g should be close to f. To ensure this, we define g as a normal with mean equal to the mode of f and covariance equal to the negative Hessian evaluated at the mode. To find the mode, we use the Newton-Raphson method. The Hessian is analytically available \citep{chan2017notes}.

The sampling algorithm follows the same sequence as in main text of the paper, however with steps $p(\boldsymbol{\psi}|\bullet)$ and $p(\eta|\bullet)$ added before sampling $\sigma^2_y|\bullet$ is sampled.

\section{Graphs}

\subsection{HS-BSTS-t Model Results}

\begin{figure}[H]
    \centering
    \includegraphics[width=\textwidth]{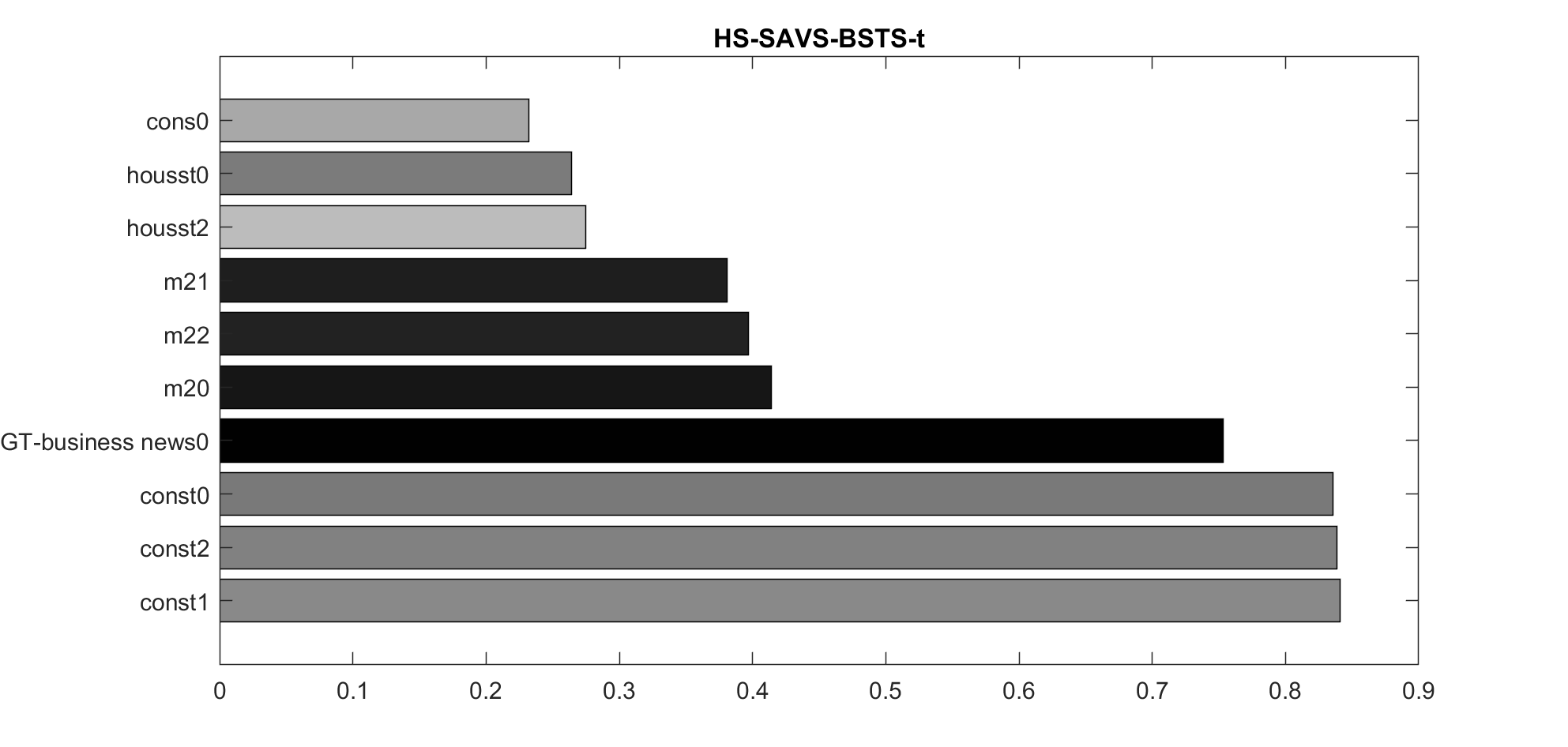}
    \caption{Posterior inclusion probabilities for the HS-SAVS-BSTS-t model.}
    \label{fig:pip_bsts_t}
\end{figure}

\begin{figure}[H]
    \centering
    \includegraphics[width=\textwidth]{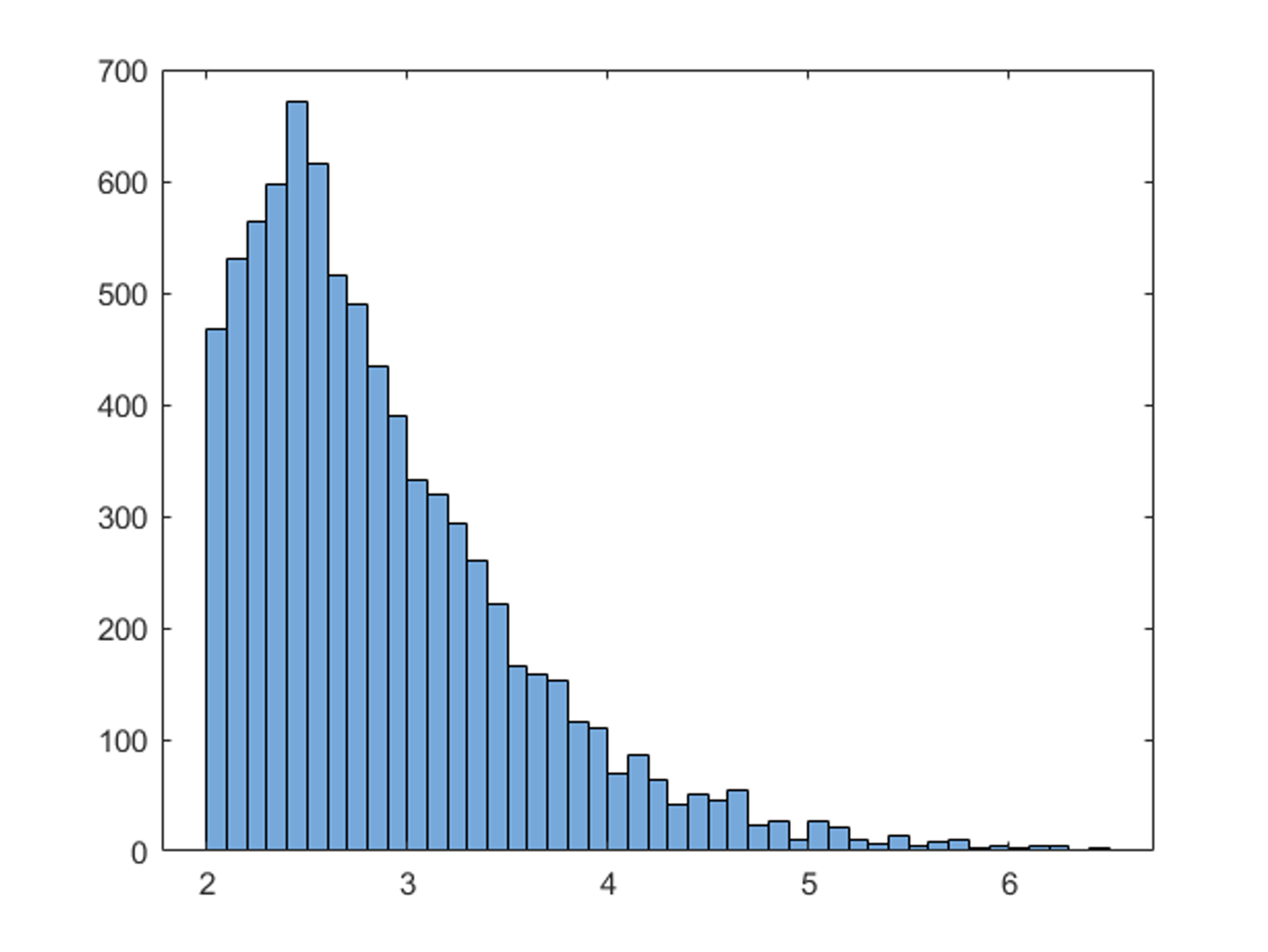}
    \caption{Posterior of the degree of freedom parameter, $\eta$.}
    \label{fig:post_eta}
\end{figure}

\begin{figure}[H]
     \centering
     \begin{subfigure}[b]{0.49\textwidth}
         \centering
         \includegraphics[width=\textwidth]{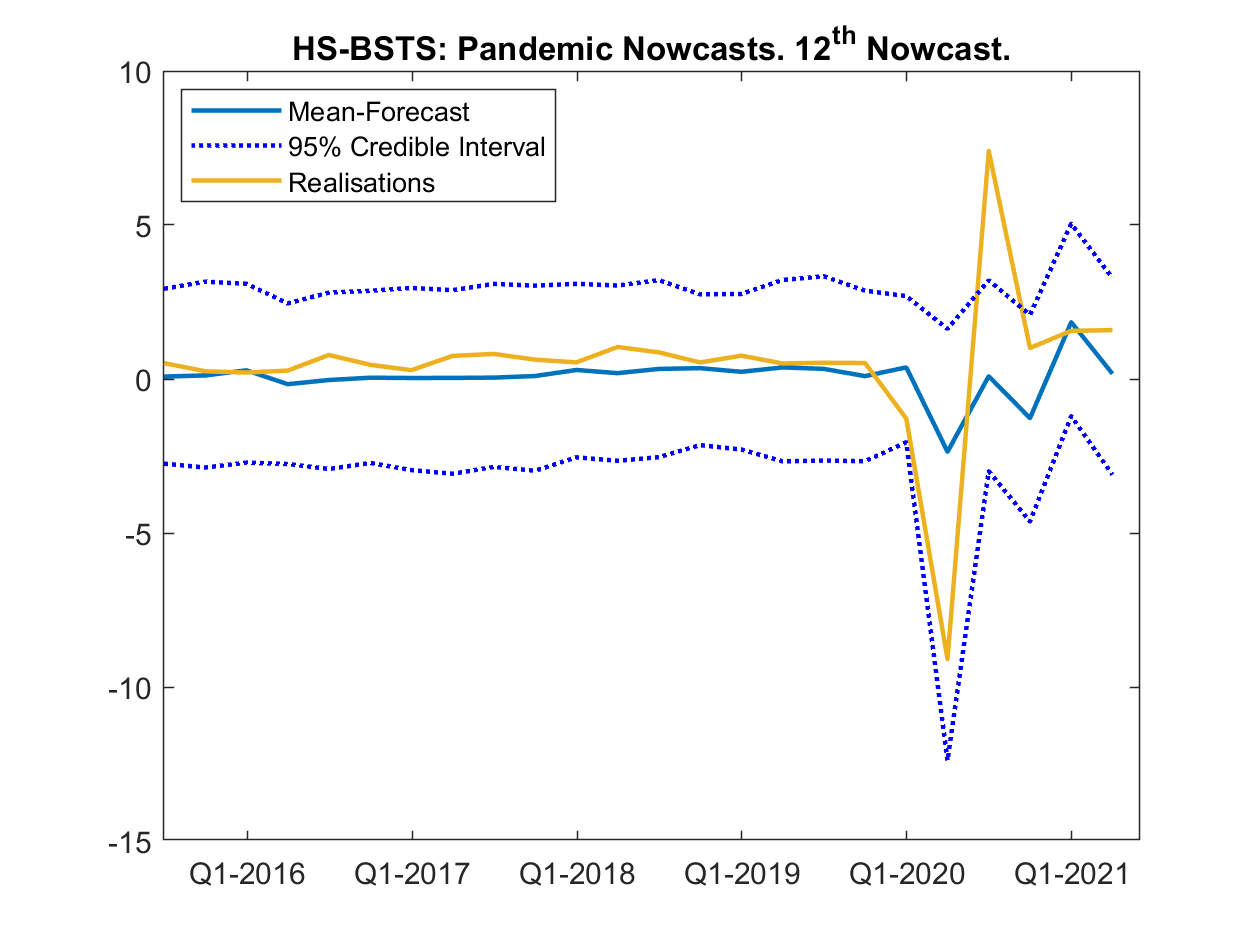}
         \label{fig:y equals x}
     \end{subfigure}
     \hfill
     \begin{subfigure}[b]{0.49\textwidth}
         \centering
         \includegraphics[width=\textwidth]{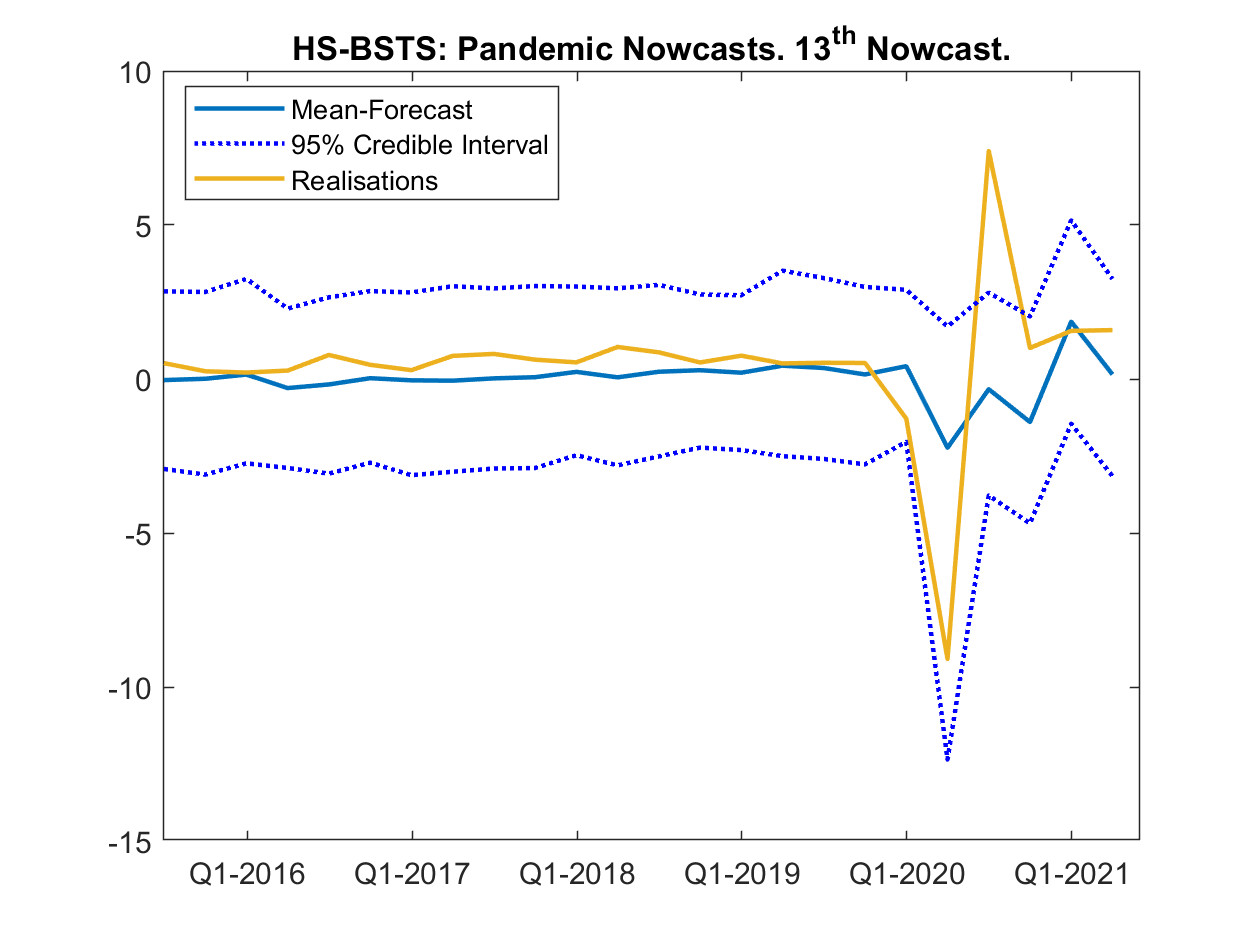}
         \label{fig:three sin x}
     \end{subfigure}
     \hfill
     \begin{subfigure}[b]{0.49\textwidth}
         \centering
         \includegraphics[width=\textwidth]{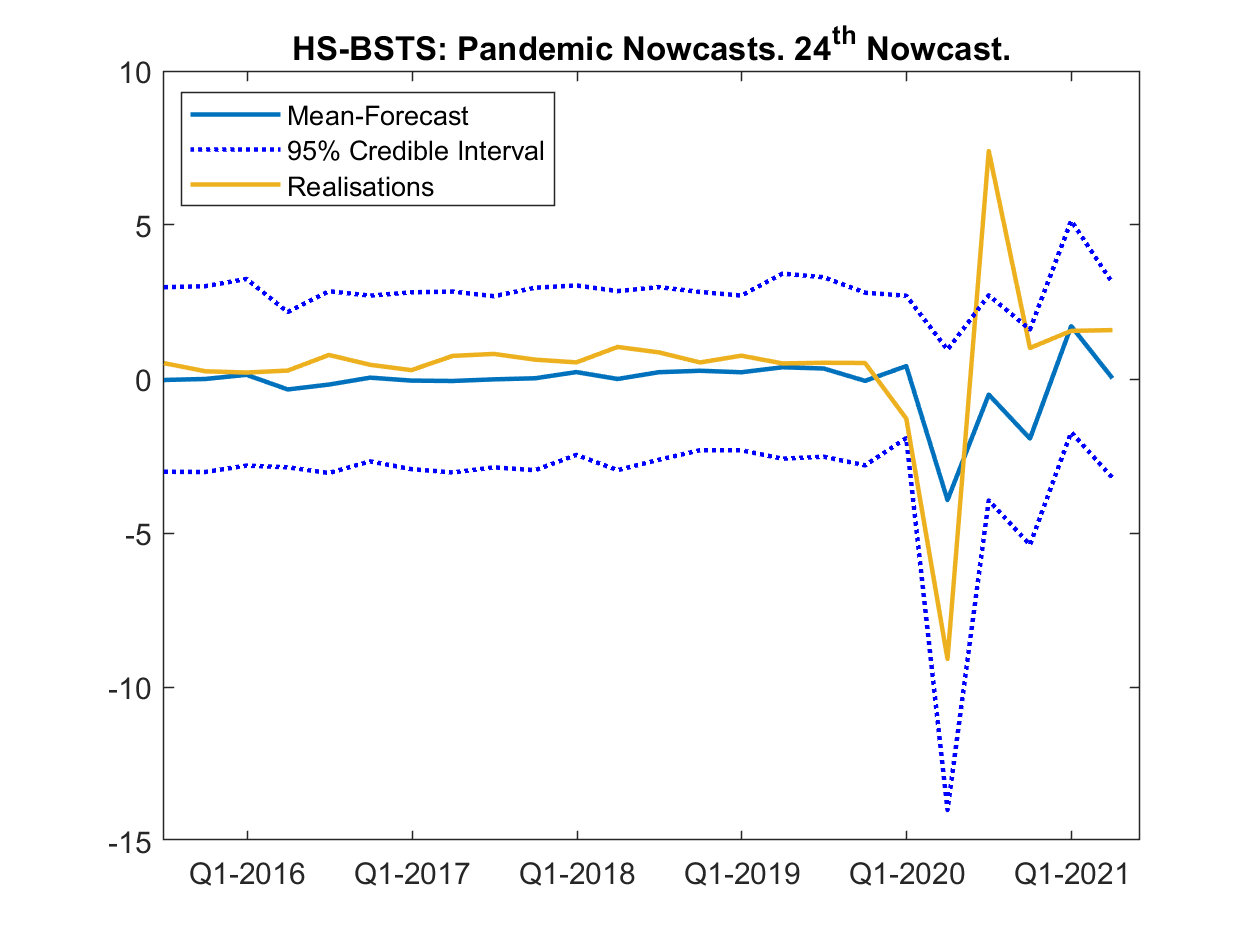}
         \label{fig:five over x}
     \end{subfigure}
     \hfill
         \begin{subfigure}[b]{0.49\textwidth}
         \centering
         \includegraphics[width=\textwidth]{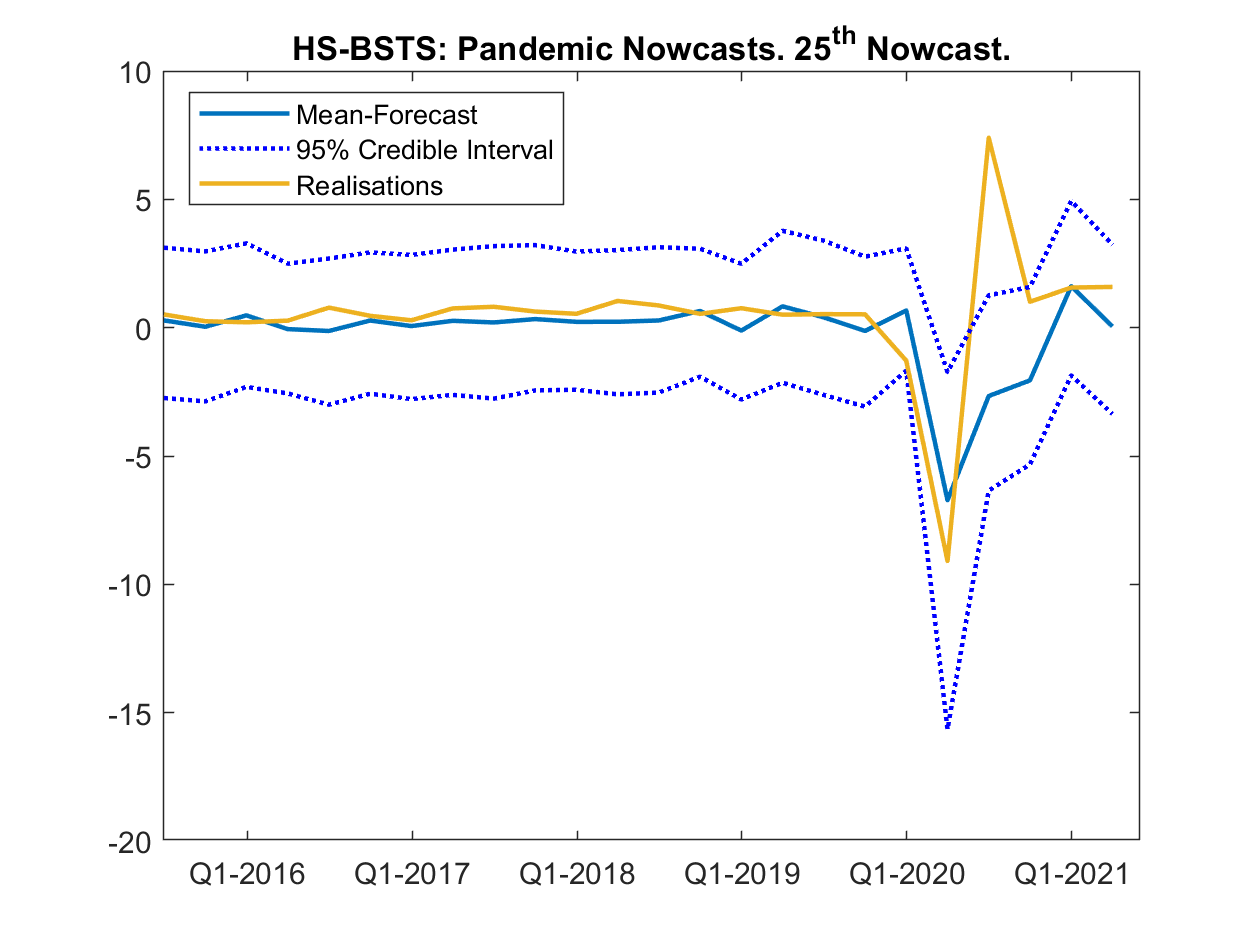}
         \label{fig:five over x}
     \end{subfigure}
        \caption{Predictive distributions for full HS-BSTS-t model for nowcast periods 12,13,24 and 25. Periods 12 and 24 are prior to Google Trends releases.}
        \label{fig:pandemic_gt}
\end{figure}

\subsection{In-Sample Results}
\begin{figure}[H]
    \centering
    \includegraphics[width=\textwidth]{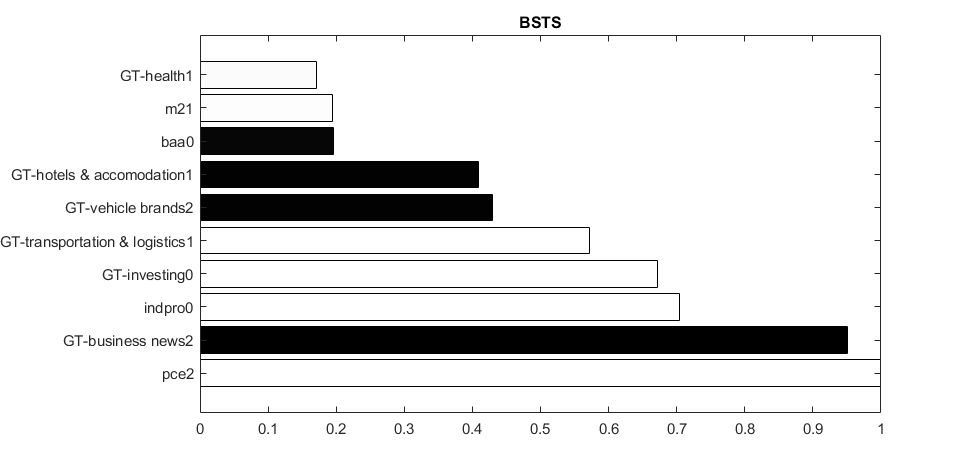}
    \caption{Posterior inclusion probabilities for the original BSTS model of \citet{scott2014predicting}.}
    \label{fig:PIPBSTS}
\end{figure}

\begin{figure}[H]
     \centering
     \begin{subfigure}[b]{0.49\textwidth}
         \centering
         \includegraphics[width=\textwidth]{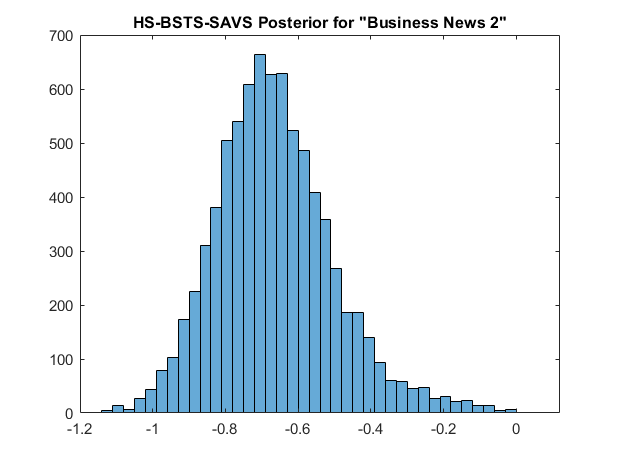}
     \end{subfigure}
     \hfill
     \begin{subfigure}[b]{0.49\textwidth}
         \centering
         \includegraphics[width=\textwidth]{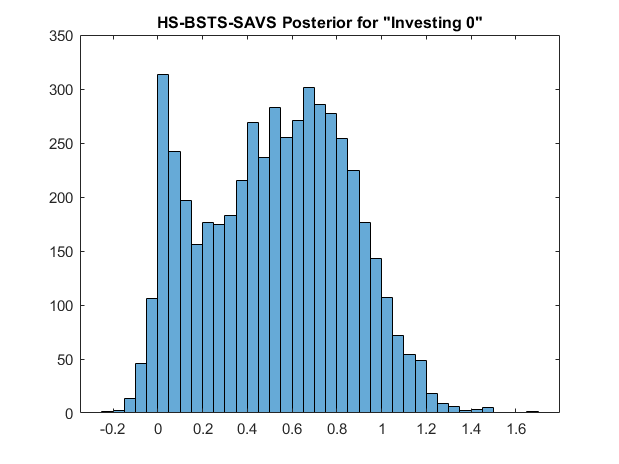}
     \end{subfigure}
    
        \caption{Posterior of the Google category 'Business news 2' (left) and topic 'Investing 0' for HS-SAVS-BSTS.}
        \label{fig:post_gt}
\end{figure}

\subsection{Google Topic/Category Plots}
\begin{figure}[H]
     \centering
     \begin{subfigure}[b]{0.49\textwidth}
         \centering
         \includegraphics[width=\textwidth]{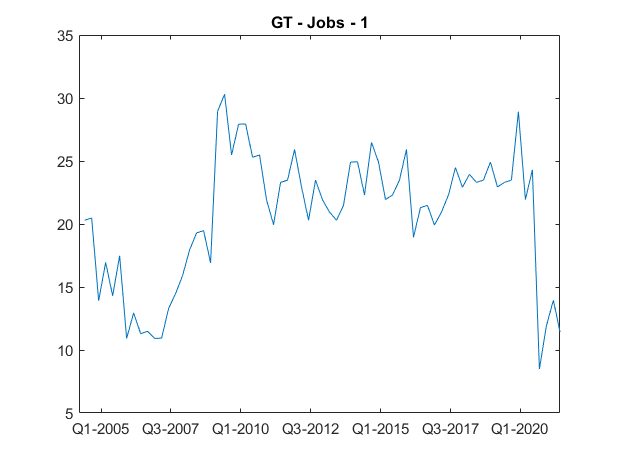}
     \end{subfigure}
     \hfill
     \begin{subfigure}[b]{0.49\textwidth}
         \centering
         \includegraphics[width=\textwidth]{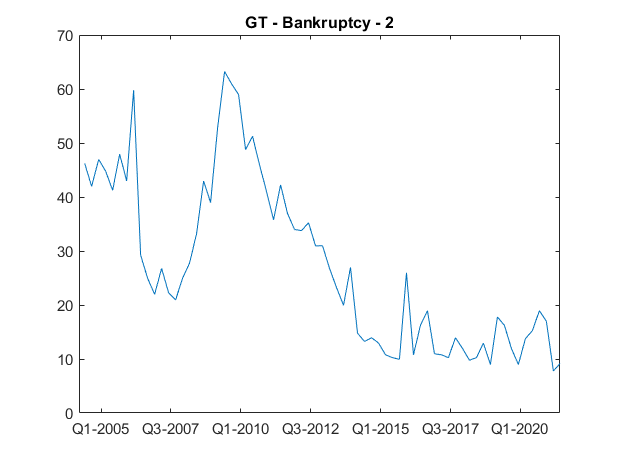}
     \end{subfigure}
     \hfill
     \begin{subfigure}[b]{0.49\textwidth}
         \centering
         \includegraphics[width=\textwidth]{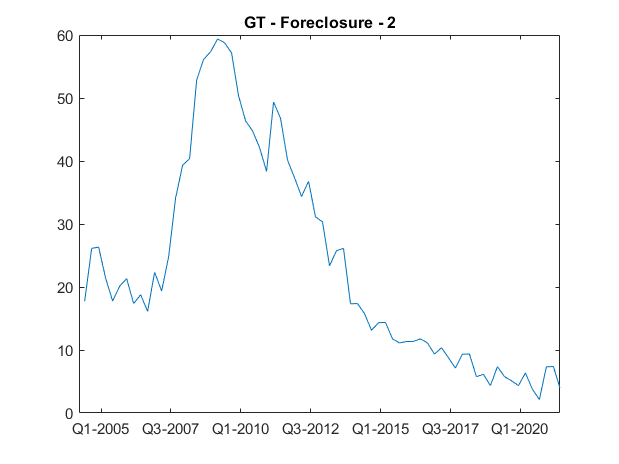}
     \end{subfigure}
     \hfill
         \begin{subfigure}[b]{0.49\textwidth}
         \centering
         \includegraphics[width=\textwidth]{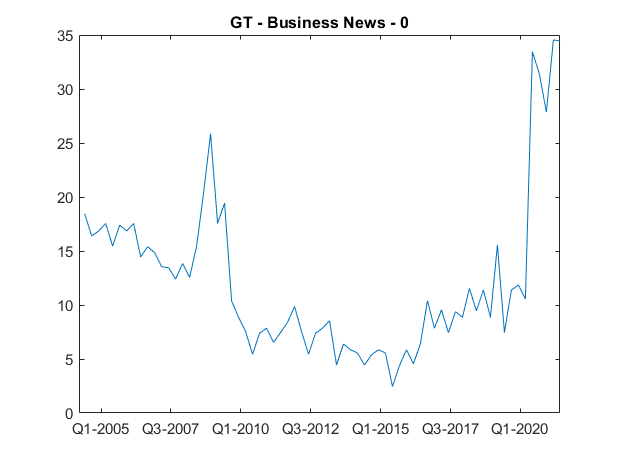}
     \end{subfigure}
  \hfill
         \begin{subfigure}[b]{0.49\textwidth}
         \centering
         \includegraphics[width=\textwidth]{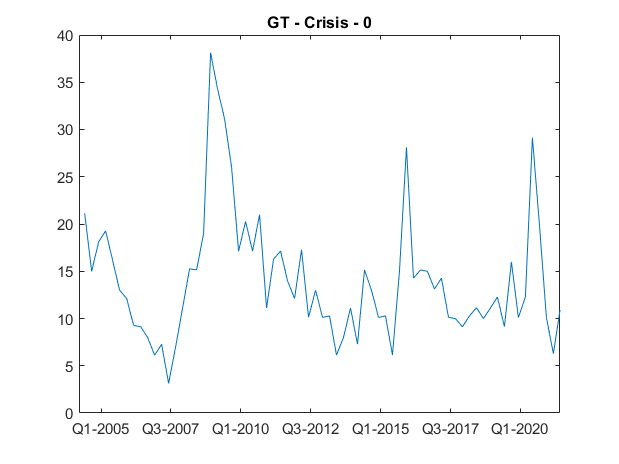}
     \end{subfigure}
  \hfill
         \begin{subfigure}[b]{0.49\textwidth}
         \centering
         \includegraphics[width=\textwidth]{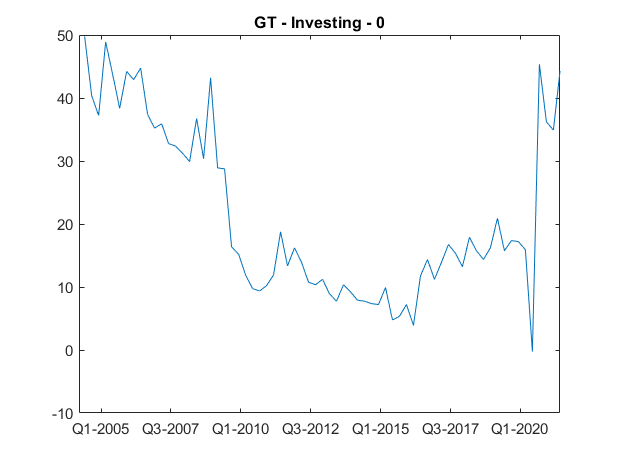}
     \end{subfigure}
        \caption{U-MIDAS transformed Google search data plots.}
        \label{fig:gtplots}
\end{figure}

\end{document}